\documentclass[aps,prd,reprint,floatfix,nofootinbib]{revtex4-2}
\usepackage{graphicx} 
\usepackage{amsmath}
\usepackage{booktabs}
\usepackage{slashed}
\usepackage{xcolor}
\usepackage[utf8]{inputenc}
\usepackage{multirow}
\usepackage{placeins}
\usepackage{hyperref}

\begin{document}

\title{Exotic \texorpdfstring{$B_s$}{Bs} mesons in the continuum \texorpdfstring{\\}{, }
from a nonperturbatively-tuned heavy quark action}

\author{R. J. Hudspith}\affiliation{Department of Physics, Carnegie Mellon University, Wean Hall 7235, Pittsburgh, PA 15213, United States}
\author{Daniel Mohler}\affiliation{GSI Helmholtzzentrum f\"ur Schwerionenforschung GmbH, Planckstraße 1, 64291 Darmstadt, Germany\\ and Institut f\"ur Kernphysik, Technische Universit\"at Darmstadt, Schlossgartenstraße 2, 64289 Darmstadt, Germany}
\author{M. Stolz}\affiliation{Institut f\"ur Kernphysik, Technische Universit\"at Darmstadt, Schlossgartenstraße 2, 64289 Darmstadt, Germany}

\begin{abstract}
In this work we predict the masses and binding energies of two $B_s$ exotic-meson candidates using Lattice QCD, namely the $B_{s0}^*$ and $B_{s1}$. We use a relativistic heavy-quark action for the valence b-quark in our simulations, tuned fully non-perturbatively by a neural network. This allows us to take the continuum limit and eliminates the largest systematic we attributed to our previous determination of these states using Lattice-NRQCD. This is the first Lattice QCD study to show that these states remain deeply bound in the continuum limit. We thoroughly benchmark our heavy-quark approach by reproducing the experimental values of the 1S hyperfine splittings of $B$ and $B_s$ mesons, as well as the mass splitting between the $B$ and $B_s$ mesons. Our final results yield binding energies with respect to the $BK$ and $B^*K$ thresholds of $-65.9(6.0)(3.0)_\text{Iso}$~MeV and $-60.6(6.6)(3.0)_\text{Iso}(1.0)_\text{GEVP}$~MeV for the $B_{s0}^*$ and $B_{s1}$ respectively.
\end{abstract}

\maketitle

\section{Introduction}

The positive-parity $D_{s0}^*(2317)$ discovered in 2003 by BaBar \cite{BaBar:2003oey} and the $D_{s1}(2460)$ discovered shortly thereafter by CLEO \cite{CLEO:2003ggt} were among the first hadrons with heavy-quarks that displayed properties not expected from simple potential models. In the picture of a heavy quark and a light quark with a mass close to the chiral limit, these two states form the $j=\frac{1}{2}$ heavy-quark multiplet, with $j$ being the total angular momentum of the light quark \cite{Bardeen:2003kt}. Their lighter-than-anticipated mass can be explained by the closeness of the $DK$ and $D^*K$ thresholds, a mechanism almost immediately suggested in \cite{vanBeveren:2003kd}. These states have been the protagonists of countless studies.\footnote{For a brief summary interested readers are refered to the PDG review on \emph{Heavy Non-$q\bar{q}$ Mesons} \cite{ParticleDataGroup:2024cfk}.} 

More generally, there is a pattern of such states \cite{vanBeveren:2003kd,Du:2017zvv}. The corresponding mesons with a single bottom quark are so far not entirely known from experiment \cite{ParticleDataGroup:2024cfk}. While the $J^P=1^+$ and $J^P=2^+$ states forming the $j=\frac{3}{2}$ states can likely be identified with the known $B_{s1}(5830)$ and $B_{s2}(5840)$, the analogous states corresponding to the exotic $D_{s0}^*(2317)$ and $D_{s1}(2460)$ in the $B_s$-sector have not yet been observed. The existence of such states has been predicted in pioneering works \cite{Bardeen:2003kt, Kolomeitsev:2003ac} after the discovery of their $D_s$ cousins, and their masses have since been predicted with varying level of rigor \cite{Guo:2006fu, Guo:2006rp,Badalian:2007yr,Colangelo:2012xi,Altenbuchinger:2013vwa,Wang:2015mxa,Cheng:2017oqh,Du:2017zvv,Alhakami:2020vil,Guo:2021rjv,Fu:2021wde,Koponen:2007nr,Gregory:2010gm,Wurtz:2015mqa,Lang:2015hza,Hudspith:2023loy}\footnote{We restrict the references provided here to those which made an attempt to estimate at least some of the uncertainties.}. Notably, this includes studies of these states based on non-perturbative QCD calculations on a Euclidean space-time lattice (Lattice QCD) \cite{Koponen:2007nr,Wurtz:2015mqa,Lang:2015hza,Hudspith:2023loy}. 

In Lattice QCD, precision studies of these states remain challenging as effects from the discretisation of the QCD action are sizable for hadrons with a heavy quark. In \cite{Lang:2015hza} these states are studied with a simple version \cite{Burch:2009az,FermilabLattice:2010rur} of an RHQ action \cite{El-Khadra:1996wdx}, where only the heavy-quark hopping parameter $\kappa_b$ has been tuned such that the spin-averaged kinetic mass of the 1S states reproduces its physical value. For this calculation a single $2+1$-flavor gauge-field ensemble with close-to-physical pion mass \cite{PACS-CS:2008bkb} was used. The pole position of the bound-state was determined through an estimate of the scattering lengths and effective range using L\"uscher's finite-volume formalism \cite{Luscher:1985dn,Luscher:1990ux}. The rough size of heavy-quark discretisation effects were estimated based on \cite{Oktay:2008ex}. More recently, these states were studied in \cite{Hudspith:2023loy} using Lattice-NRQCD. While this approach allowed for a combined extrapolation of the observed binding energies to the physical quark-mass and infinite-volume limits, there is no formal continuum limit in Lattice-NRQCD. The discretisation uncertainty was estimated by performing Lattice-NRQCD calculations at two different lattice spacings within the region where one has hope that the framework is applicable; the resulting uncertainty from the use of Lattice-NRQCD was the dominant systematic attributed to the binding energies. 

Given the current mass predictions, the as-of-now experimentally unobserved $j=\frac{1}{2}$ $B_s$ states are expected to decay into S-wave states by either emitting a photon or a $\pi^0$ \cite{Bardeen:2003kt}. In principle, these hadrons should be observable by the LHCb or Belle~II experiments \footnote{During the writing of this manuscript members of the LHCb alerted us of preliminary results presented in a talk at ICHEP regarding the discovery of the $B_{s0}^*$.}.

In the following, we report on a determination of the masses of the lowest positive-parity states with $J^P=0^+$ and $1^+$ of the $B_s$ from Lattice QCD, in which most systematic uncertainties are controlled and the binding energies (relative to the $B^{(*)}K$ thresholds) can be determined at physical hadron masses and in the combined continuum and infinite-volume limits, enabling a fully-controlled prediction for experiment. To this end, we employ a relativistic heavy quark (RHQ) action in the form of \cite{Aoki:2003dg}, which we tune nonperturbatively using the method described in \cite{Hudspith:2021iqu}. The paper is organized as follows: In the next section our heavy-quark action and non-perturbative tuning are described. In section \ref{sec:tests} we test this action on the 1S hyperfine splittings in the $B$ and $B_s$ meson spectrum. as well as the $B_s-B$ mass-splitting. These quantities are known to be sensitive to both the heavy-quark mass and to heavy-quark discretisation effects. Our results reproduce the experimental splittings to an appreciable accuracy upon taking the continuum limit. With the confidence gained from these precision tests we calculate the masses of the lowest $J^P=0^+$ and $1^+$ $B_s$ mesons, and the prediction of their masses constitutes the main result of this paper. We conclude with some brief remarks and compare our determination to the literature.

\section{Methodology}

For the b-quark, we use the Relativistic Heavy Quark (RHQ) action from \cite{Aoki:2003dg} characterised by 5 free parameters $\kappa,r_s,\nu,c_E,$ and $c_B$, with the conventional $r_t=1$. This action is argued to have discretisation effects at $O((a\Lambda_\text{QCD})^2)$. We will use the same nonperturbative tuning setup for this action as we did for charm in \cite{Hudspith:2021iqu}, now extending it to the regime of the bottom quark, with the target spectrum being that of low-lying bottomonium (the $\eta_b,\Upsilon,\chi_{b0},h_b,$ and $\chi_{b1}$) at the light and strange quark $\text{SU}(3)_f$-symmetric point. This means $B$ and $B_s$ meson masses and splittings can serve as a benchmark of the quality of our heavy-quark tuning, as they are not part of the parameter tuning. The 1S hyperfine and 1P-1S splittings in the bottomonium spectrum should be sensitive to changes in $c_E$ and $c_B$ as was seen from our nonperturbative NRQCD tuning in \cite{Hudspith:2023loy}, and so our tuning should be able to reliably ascertain differences between these parameters. As the focus of this work is on P-wave excited states of $B_s$ mesons, it is our expectation that accurately tuning to these in the bottomonia sector will carry over to the heavy-light meson sector.

\subsection{Quark sources and propagators}\label{sec:propsandsources}

We found inverting heavy quark propagators at the bottom mass quite challenging in comparison to our previous work with charm, as we discuss in App.~\ref{app:hopping}. In the end our solution to the problem was to trade a residual stopping condition for a truncation criterion in the Hopping Parameter Expansion (HPE). For small $\kappa$ values (as our tuning gives for b-quarks) the number of iterations required is low, but for finer boxes the number of fixed CG iterations or HPE iterations must grow. At large times we can find ourselves in a situation where the HPE series hasn't saturated and the effective mass of, say, our B-meson will grow, as in Fig.~\ref{fig:CGHPE} of App.~\ref{app:hopping}. Such behaviour is unphysical and can make robust plateau identification challenging.

For all of our quark propagators we used Coulomb gauge fixed\footnote{fixed to a precision of $10^{-14}$ using the FACG algorithm of \cite{Hudspith:2014oja}} wall sources with a Gaussian profile applied at the source \cite{Davies:1994mp} and point-like sinks. This differs from the previous setup \cite{Hudspith:2023loy,Hudspith:2020tdf}, which used gauge fixed wall sources with Gaussian (box) sink smearing.

The smearing is used to reduce excited-state contamination, as gauge-fixed wall sources with point-like sinks typically suffer from higher states contributing with negative amplitudes (i.e.~effective masses that approach a plateau from below). Performing the smearing at the source this time was merely a technical consideration, as the propagators and contractions in this calculation were computed on-the-fly rather than stored and post-processed. We observe that gauge fixed walls have much better statistical resolution than stochastic sources for our heavy-light mesons.

Generically, the meson operators we will consider throughout this work are the site-local,
\begin{equation}
O_\Gamma = ( \bar\chi \Gamma \psi),
\end{equation}
with quark flavors $\chi$ and $\psi$. For the bottomonia-based RHQ parameter tuning we determine ground states with $\Gamma=\gamma_5,\gamma_i,I,\gamma_i\gamma_5,\gamma_i\gamma_j$ and $\chi=\psi=b$. Throughout this work, we will only use the quark-connected Wick contraction for mesons. Noting that for bottomonia disconnected diagram contributions will be heavily mass-suppressed, and of course for our heavy-light mesons these diagrams are absent. For the $B$ and $B_s$ hyperfine splittings we will consider $\Gamma=\gamma_5,\gamma_t\gamma_5$ at the source and sink for the pseudoscalar and $\Gamma=\gamma_i,\gamma_i\gamma_t$ for the vector (with $\chi=b,\psi=u/d,s$). For the exotic $B_s$-mesons, the $B_{s0}^*$ and $B_{s1}$, we will consider the $\Gamma=I$ and $\Gamma=\gamma_i\gamma_5$ interpolators respectively (with $\chi=b,\psi=s$).

In Tab.~\ref{tab:acceptable_pars} we list the minimum acceptable number (which is what we use) of HPE iterations for the heavy quarks, determined from minimising the cost of runs while having a reliable plateau for the $\eta_b$. We also list the source-smearing radius (squared) we used and we set this to be at fixed $r\approx 0.56 \text{ fm}$, determined to yield long plateaus for the $T_{bb}$ (which will be the focus of a future publication) and our excited $B_{s0}^*$ and $B_{s1}$. Unfortunately, this turns out to be a somewhat sub-optimal choice for the B and $B_s$ mesons. This \textit{optimal} smearing radius for our RHQ parameters is larger than the one for lattice NRQCD we used in \cite{Hudspith:2023loy} ($r\approx 0.41 \text{ fm}$), perhaps indicative of the need to suppress excited states for the RHQ.

\begin{table}[tb]
\centering
\begin{tabular}{c|cc|cc}
\toprule
Name & $\beta$ & $a^{-1}$ [GeV] & $N_{\text{HPE}}$ & $(r/a)^2$ \\
\hline
A653 & 3.34 & 1.987(20) & 64 & 32 \\
U103 & 3.40 & 2.285(28) & 94 & 42 \\
B450 & 3.46 & 2.585(33) & 88 & 54 \\
H200 & 3.55 & 3.071(36) & 96 & 76 \\
N300 & 3.70 & 3.962(45) & 142 & 126 \\
J500 & 3.85 & 5.047(58) & 200 & 198 \\
\botrule
\end{tabular}
\caption{Number of HPE iterations used and source smearing radius applied.}\label{tab:acceptable_pars}
\end{table}

\subsection{Heavy-Quark Tuning}

\begin{figure*}[bt]
\center
\includegraphics[scale=0.26]{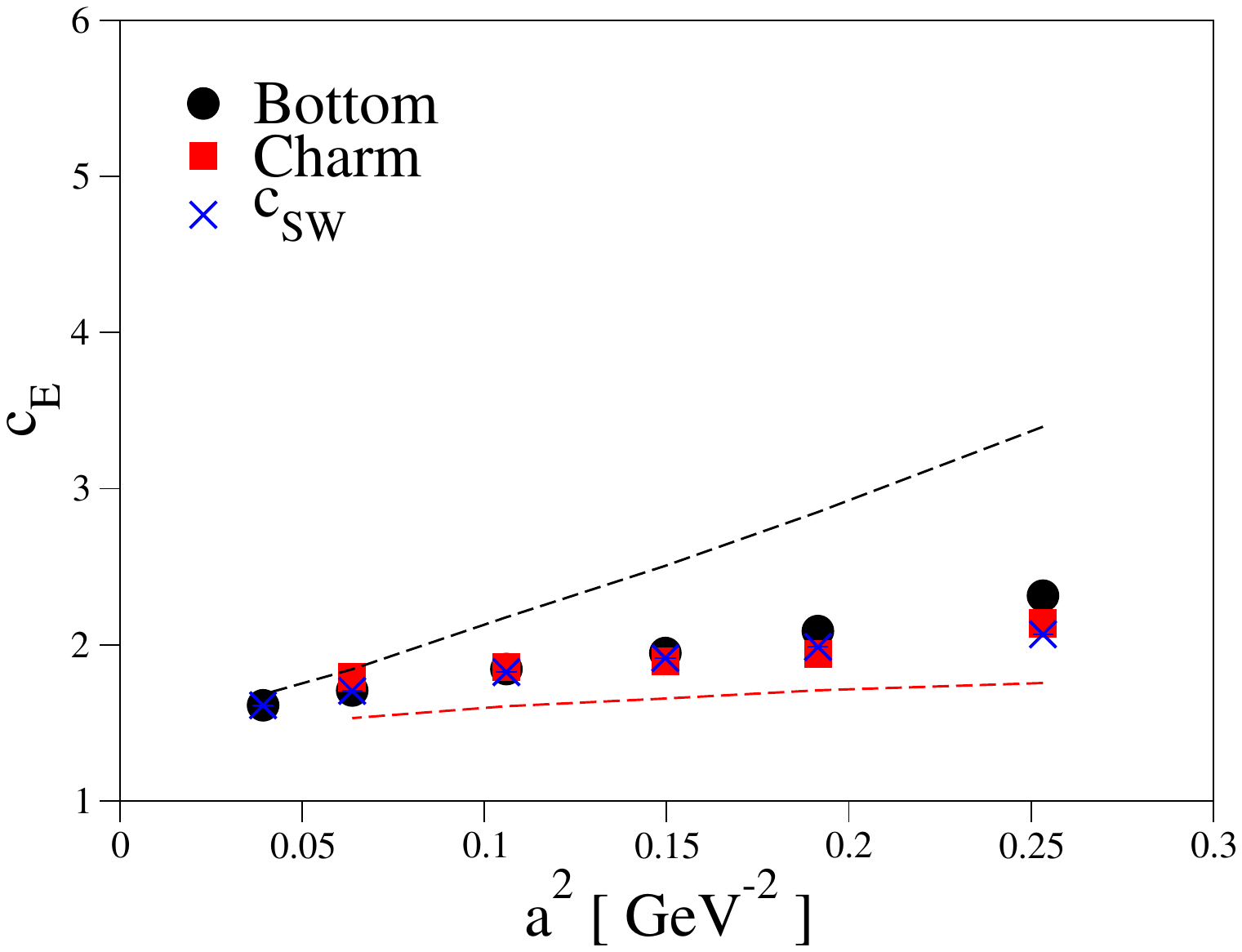}
\includegraphics[scale=0.26]{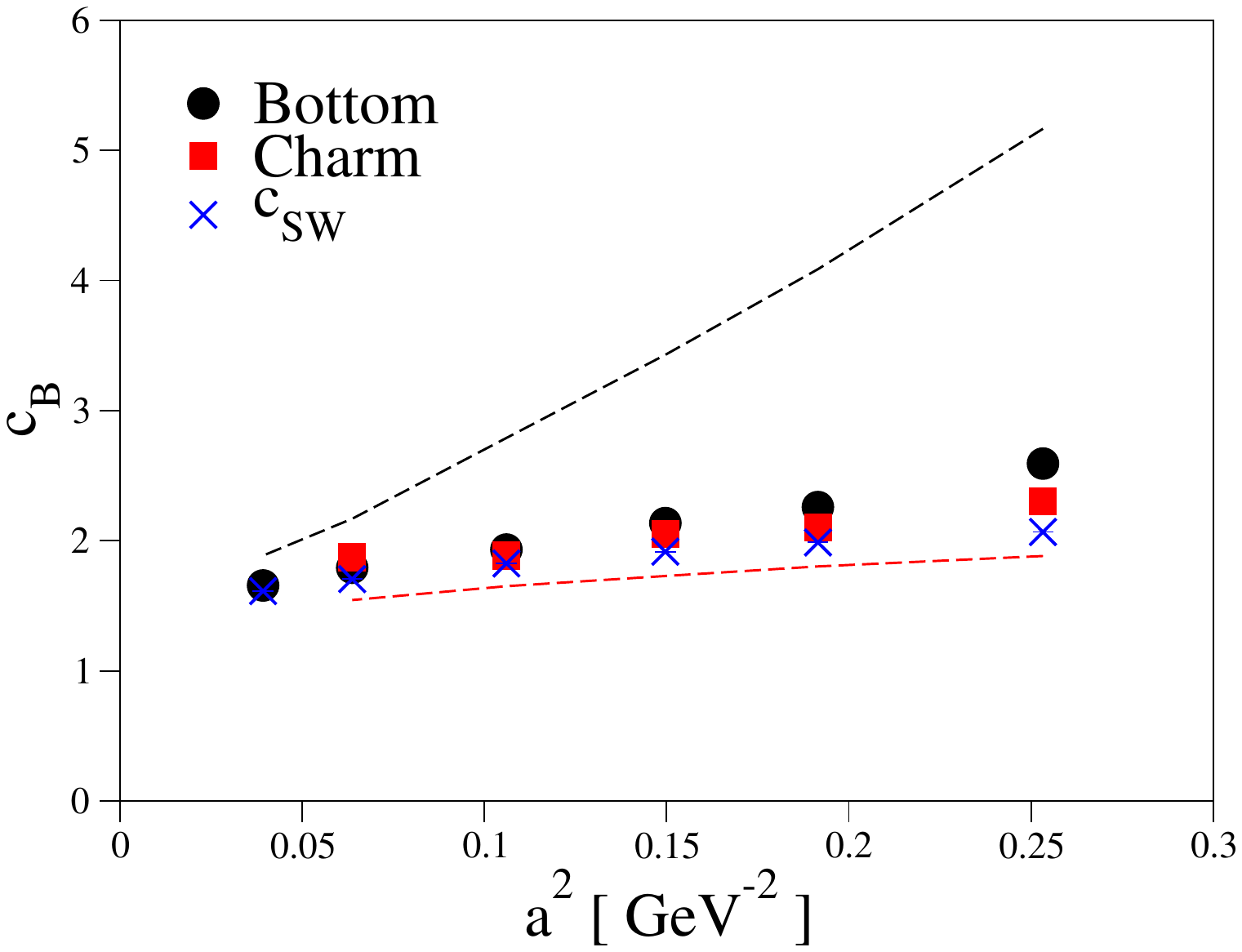}\\
\includegraphics[scale=0.26]{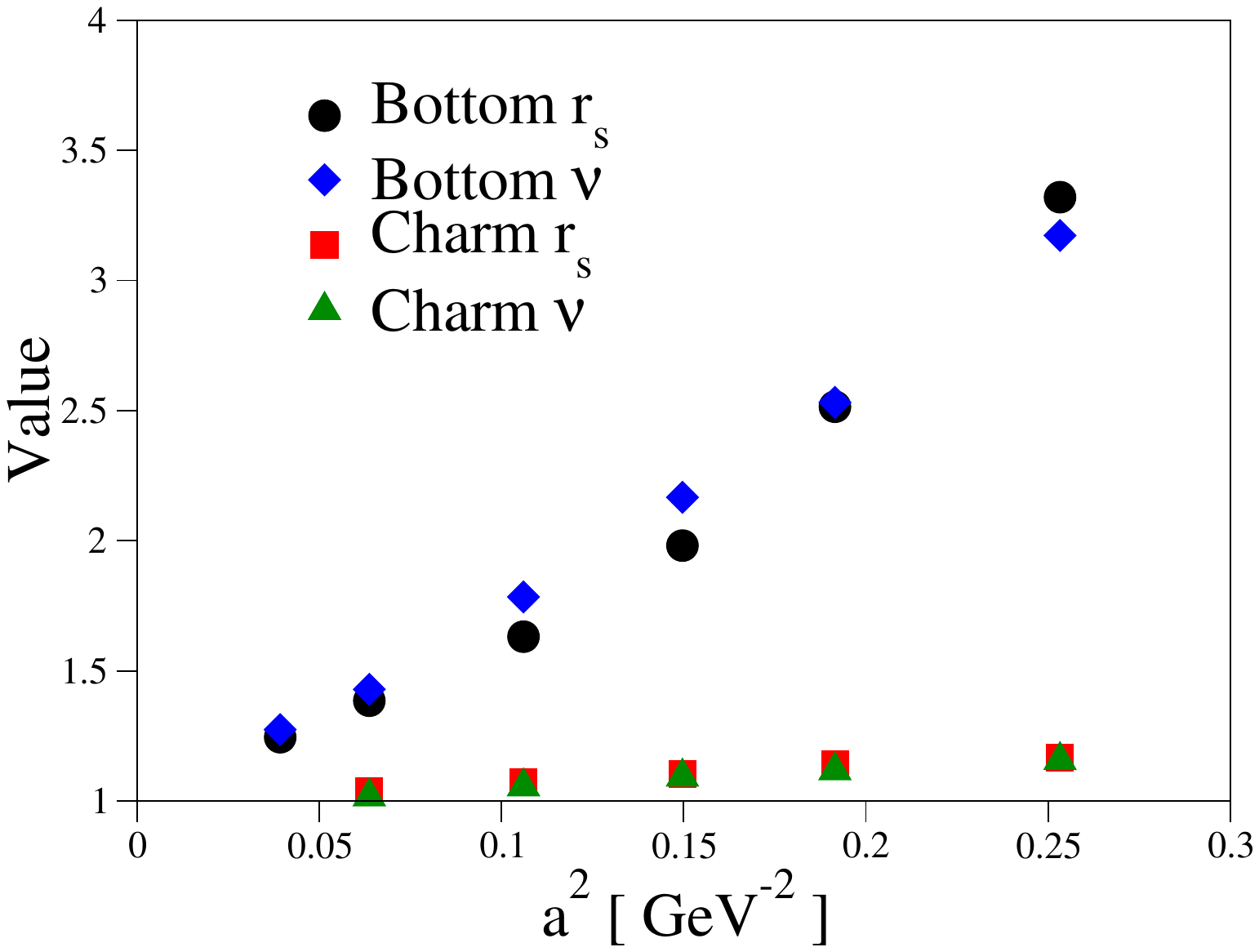}
\includegraphics[scale=0.26]{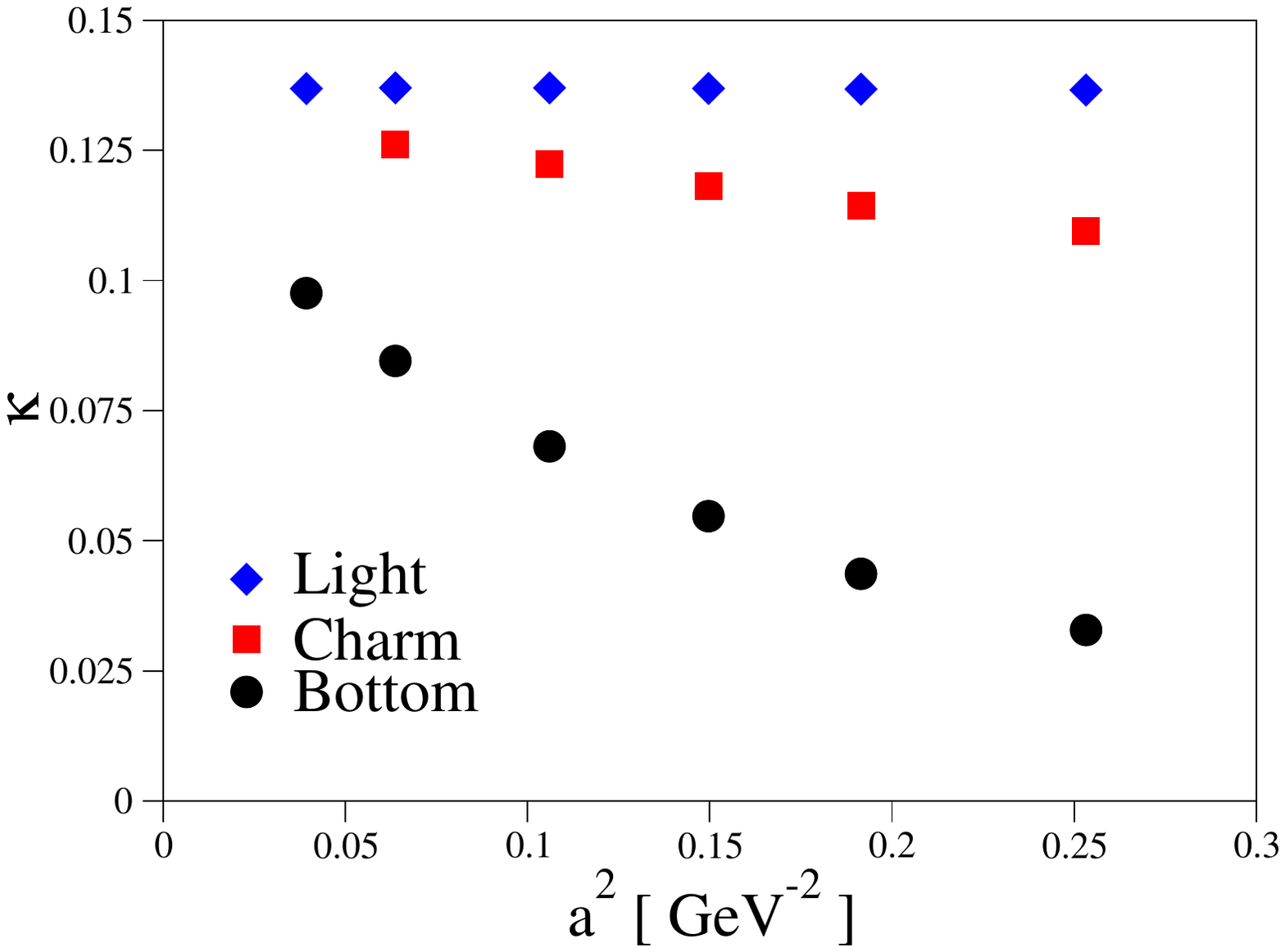}
\caption{Our b-quark tuning parameters compared to those we obtained for the charm quark in \cite{Hudspith:2021iqu} from lattices with the same $\beta$-values. The top row shows $c_E$ (left) and $c_B$ (right) dashed lines indicate the tree-level tadpole-improved perturbative expressions $c_E=\frac{1+\nu}{2u_0^3}$ and $c_B=\frac{\nu}{u_0^3}$. The bottom row shows $r_s$ and $\nu$ vs. $a^2$ (left) and $\kappa$ vs $a^2$ (right).}\label{fig:comp_tuning}
\end{figure*}

We use randomly-drawn RHQ action parameters, compute the low-lying bottomonium spectrum ($\eta_b,$ $\Upsilon,$$\chi_{b0},$$\chi_{b1}$,$h_b$), and measure the spin-averaged $\eta_b$ and $\Upsilon$ speed of light $c^2$. We train a neural network to infer the pattern between the given parameters and the computed states for many runs. 
We then feed the network the physical masses (combined with $c^2=1$) and let it predict our optimal parameters at fixed $\beta$. Here, we will only work with ensembles at the $\text{SU}(3)_f$ flavor-symmetric point on small boxes as the bottomonium-spectrum should not be particularly sensitive to light-quark or finite-volume effects. In contrast to our tuning for charmonium \cite{Hudspith:2021iqu}, we used 100 distinct sets of randomly-drawn parameters for each ensemble and utilized two hidden layers in our neural network.

Our methodology depends on the lattice scale for conversion to physical units (as is the case for most heavy-quark tunings that match a target mass or masses); we will use the values from \cite{Bruno:2016plf} (and the estimate for the $\beta=3.34$ and $\beta=3.85$ lattices, A653 and J500, from \cite{Chao:2021tvp} and \cite{Chao:2022xzg}). As newer lattice spacings from different observables are computed the existing set of runs can, in principle, be reused to provide new parameters if the predictions lie somewhere within the training set. In taking the continuum limit, the matching scales and RHQ parameters must be used self-consistently, and so we list the values we used in Tab.~\ref{tab:acceptable_pars}.

In Tab.~\ref{tab:bparams_v2} we provide the parameters and the 1$\sigma$-intervals from different neural network predictions. It is clear that the parameters $r_s$ and $\nu$ become very large with very heavy masses and coarse lattice spacings.  Presumably this is the neural network attempting to minimise the sizeable cutoff effects. Our optimal parameters for $r_s$ and $\nu$ are larger for our finest ensemble than for the charm quark with our coarsest ensemble, and these parameters must go to 1 in the continuum. This indicates that controlling our discretisation effects will be the most significant hurdle faced in this work, and justifies our use of several lattice spacings.  

\begin{table}[tb]
\centering
\begin{tabular}{c|ccccc}
\toprule
Name & $\kappa$ & $c_E$ & $c_B$ & $r_s$ & $\nu$ \\
\hline
A653 & 0.03284(2)  & 2.315(5) & 2.593(85) & 3.320(8) & 3.173(14) \\
U103 & 0.04363(4)  & 2.088(6) & 2.258(10) & 2.515(5) & 2.532(11) \\
B450 & 0.05473(2)  & 1.947(3) & 2.135(6)  & 1.981(2) & 2.167(6) \\
H200 & 0.06813(14) & 1.845(3) & 1.932(5)  & 1.631(4) & 1.784(3) \\
N300 & 0.08453(4)  & 1.707(1) & 1.792(3)  & 1.385(2) & 1.429(1) \\
J500 & 0.09756(3)  & 1.613(2) & 1.656(4)  & 1.245(2) & 1.274(4) \\
\botrule
\end{tabular}
\caption{RHQ action parameters for the b-quark, with associated variations. 
Here we list the tuning parameters for the ensemble J500 although it is not used in our current study of the exotic $B_s$-mesons.
}\label{tab:bparams_v2}
\end{table}

\begin{figure}[tb]
\center
\includegraphics[scale=0.26]{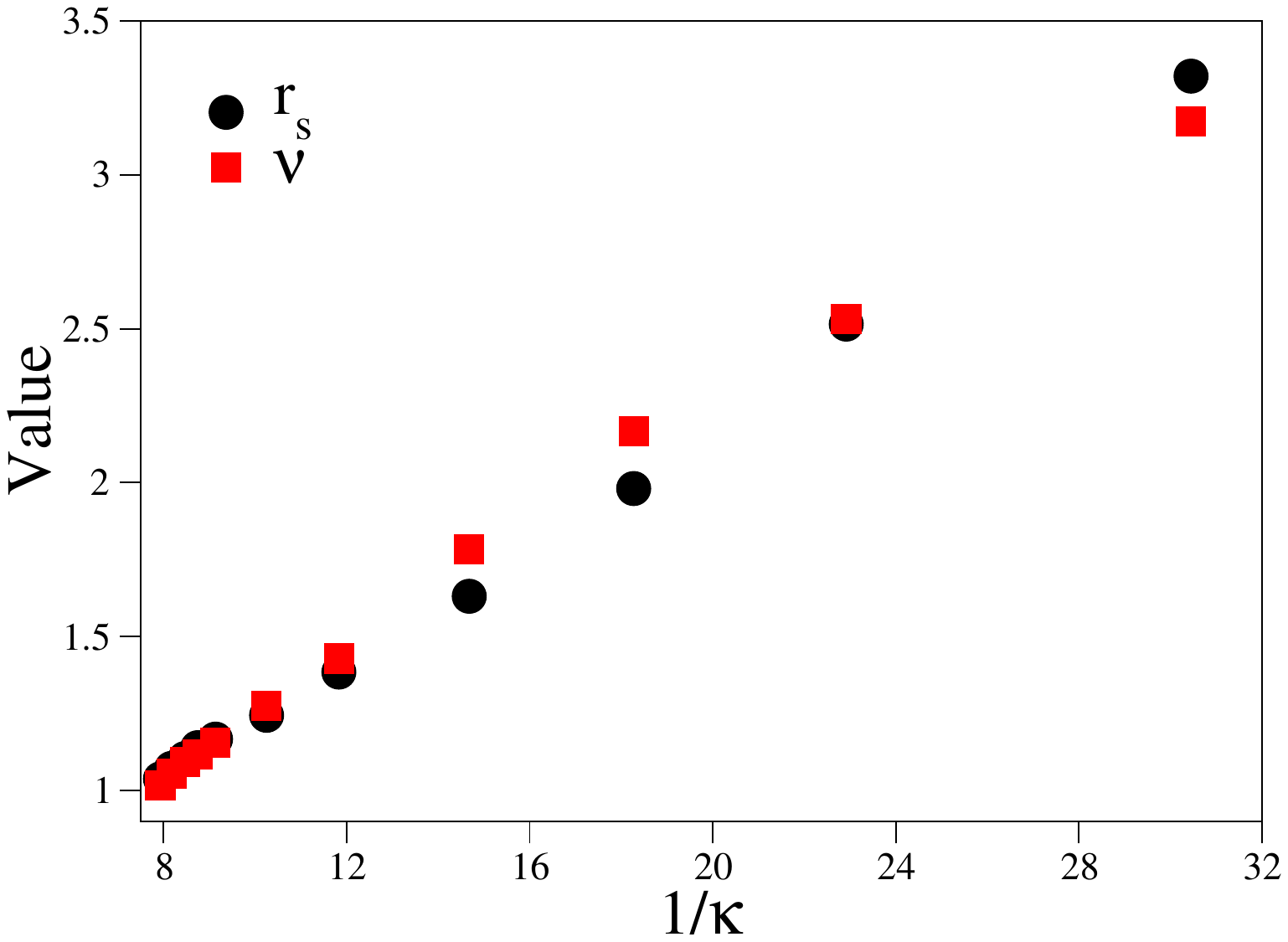}
\caption{A comparison of our values for $r_s$ and $\nu$ against our tuned heavy-quark $1/\kappa$ for both charm (from \cite{Hudspith:2021iqu}) and bottom.}\label{fig:comp_rsnuinv}
\end{figure}

A comparison of our tuning parameters for charm and bottom quarks can be found in Fig.~\ref{fig:comp_tuning}. For both charm and bottom we have $c_E\approx c_{SW}$ (the non-perturbatively tuned clover term from \cite{Bulava:2013cta}), and we always have the hierarchy $c_B>c_E$ as might be expected from non-relativistic QCD or tree-level perturbation theory. For both charm and bottom quarks their values are comparable, with the bottom leading to slightly larger values than the charm at the same $\beta$ (particularly for $c_B$). This difference is on the order of 12/7\% (bottom/charm) from $c_{SW}$ for our coarsest lattice and still measurable at our finest.

Our tuning is therefore never compatible with a universal $c_E=c_B=c_P$ as suggested by \cite{Lin:2006ur,Christ:2006us}, presumably due to $(ap)^2$ and higher terms. As one may expect, our tuning parameters are sensitive to cut-off effects; our values are dependent on a specific definition of the lattice spacing. For example, if our coarsest lattice spacing were smaller than the value used here, the coefficients $\kappa$, $c_E$, and $c_B$ would increase, $\nu$ decrease, and $r_s$ would only mildly depend on this variation.
Provided we can take a continuum limit and are self-consistent in our use of the scales, the parameters are just a reflection of this choice. However, if the same CLS ensembles are used with the scales of \cite{Bruno:2016plf} for a similar RHQ action and the tadpole-improved perturbative values of $c_E$ and $c_B$ are chosen then one would expect strong cutoff effects in spin splittings.

Our nonperturbatively tuned clover parameters $c_E$ and $c_B$ are in conflict with the mean-field tadpole-improved expectations \cite{Chen:2000ej} of $c_E = \frac{1+\nu}{2u_0^3}$ and $c_B=\frac{\nu}{u_0^3}$, albeit they are for an action where $r_s=\nu$, a relation that holds for our tuning at worst at the 10\% level for bottom and 2\% for charm. These predictions are indicated by dashed lines in Fig.~\ref{fig:comp_tuning}. We note that the rapidly-rising values of $r_s$ and $\nu$ do not strongly impact our nonperturbatively determined $c_E$ and $c_B$ and our training data always includes the variations $c_E=c_B$ as well as $c_B<c_E$.

We find that while $\nu < r_s$ holds for charm, for the majority of our ensembles the opposite is the case for bottom, and there is a noticeable curvature for $r_s$ in $a^2$. Plotting both our bottom and charm tuning parameters for $r_s$ and $\nu$ vs $\kappa^{-1}$ in Fig.~\ref{fig:comp_rsnuinv}, we see strong linear dependence, which is to be expected as $\kappa^{-1} \propto am$. There appears some opposite-signed curvature in $\nu$ at higher orders in the mass that causes this inversion in hierarchy in the tuned parameters between charm and bottom.

Finally, we summarise guidelines for performing a comparable heavy-quark tuning. As a first guess, we recommend tuning kappa and $\nu$ very roughly to get the overall $\eta_b$-mass correct with $c^2\approx1$. For this, one can set $c_E=c_B=c_{SW}$ (where $c_{SW}$ could just be the tadpole-improved value) and $r_s=\nu$, this can be done for a couple of runs and will persist in the training set. Then all of the parameters can be varied with somewhat wide Gaussians (including the tadpole-improved perturbative expectations of $c_E$ and $c_B$) initially; these widths can then be shrunk on further passes. Using the predictions of earlier sets of runs in the final training set can greatly speed up convergence.

\subsection{Determination of the spectrum}

\begin{figure}[tb]
    \centering
    \includegraphics[width=0.45\textwidth]{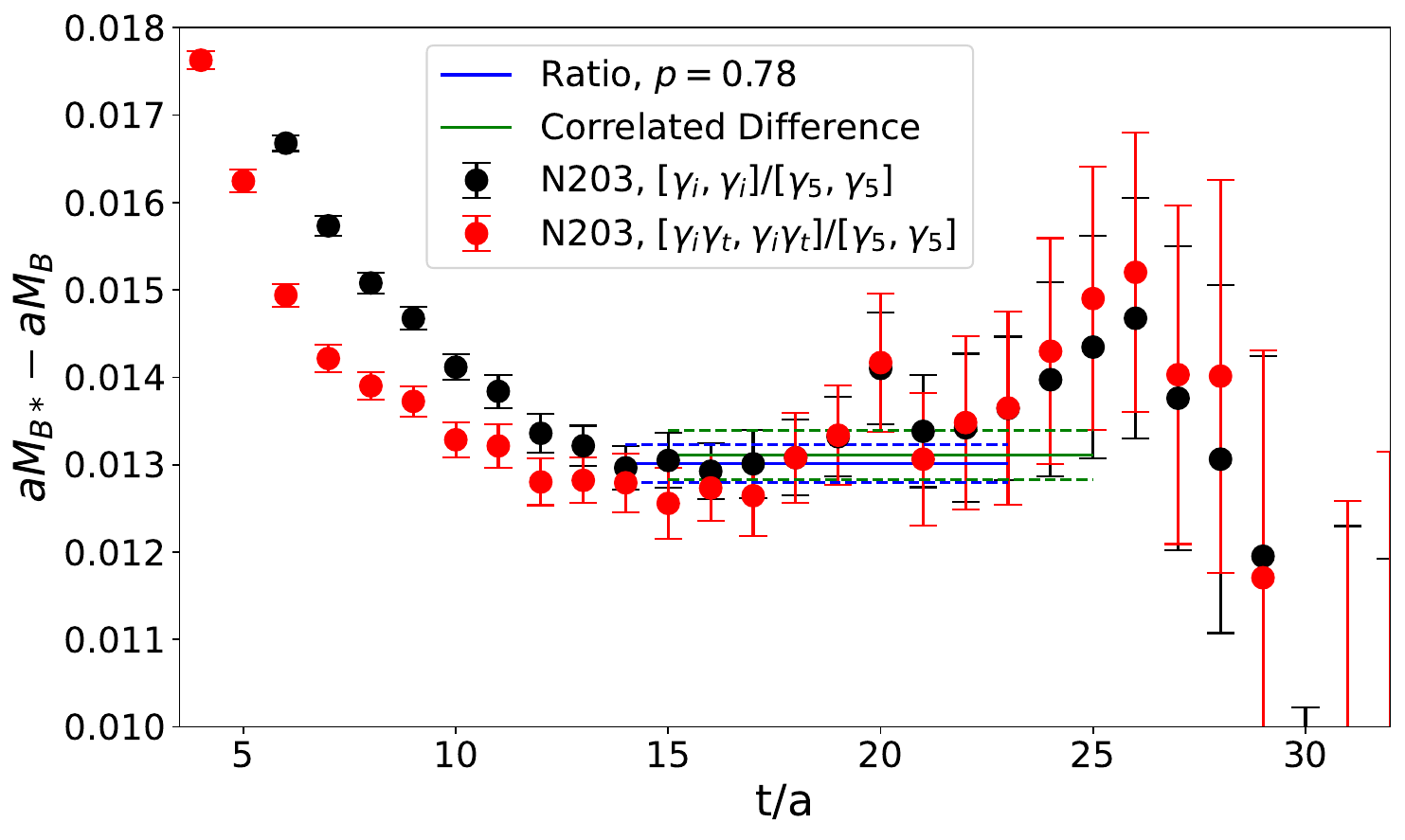}
    \includegraphics[width=0.45\textwidth]{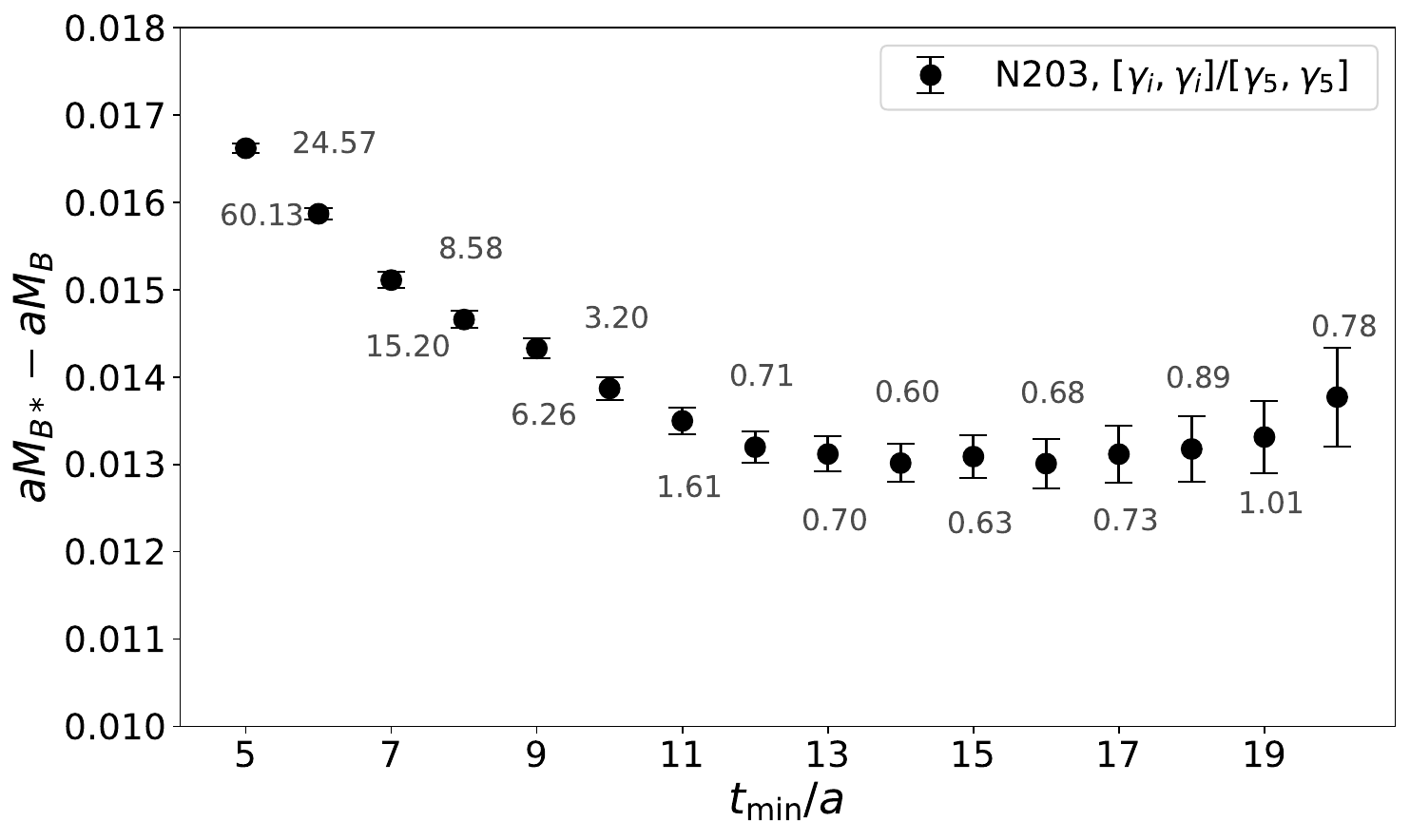}
\caption{The upper pane shows example effective masses from different interpolating field operators for the ratio-based determination of the B-meson hyperfine splitting on N203. The blue and green central values and uncertainty bands show our preferred determinations from either fits to a ratio of correlators or from correlated differences obtained through seperate fits of the two states. For an explanation of the methodology please refer to the main text. The lower shows the $t_{min}$-dependence of single-exponential fits of one of these correlation functions and the associated $\chi^2$ per d.o.f. of the fit.}
\label{fig:hyperfine_B_Meff_Stab}
\end{figure}

The crucial analysis step for the main results section in this work consists of the determination of the ground states from correlation functions employing the interpolating fields and sources described in Sec.~\ref{sec:propsandsources}. Here the procedure used to arrive at reliable ground-state masses is briefly outlined.

Energy differences have been determined with two methods: In the first, the single masses entering the determined mass difference (i.e. mass splitting or binding energy) are obtained by separate single-exponential fits to the corresponding correlators. Here and in the following, bootstrap resampling with 2000 resamples is employed throughout. The energy differences are 
obtained by subtracting the determined masses sample-by-sample and determining the central value and bootstrap uncertainty for the final observable. We refer to this method as \emph{correlated differences}.

The second method consists of taking ratios of correlation functions and performing single-exponential fits to these ratios to directly determine the corresponding energy difference. We refer to this method as \emph{ratio fits}, and it is expected that some systematics will cancel in such a procedure provided the correlation functions are similar. The two methods should agree and will both be used to form our model average. 

To determine a reasonable upper fit range for both kinds of fits we inspect effective-mass plots, keeping in mind that both the use of the HPE and the open boundary conditions as well as the statistical noise, limit the maximum reasonable fit range. An example of effective masses from the ratio method for two out of four available ratios on ensemble N203 is shown in the upper pane of Fig. \ref{fig:hyperfine_B_Meff_Stab}. Once this choice has been made, we vary the lower fit range $t_{min}$ and monitor the stability of the result and the associated goodness of fit quantified by the $\chi^2$/d.o.f. The lower pane of Fig. \ref{fig:hyperfine_B_Meff_Stab} shows such a stability plot for one of the correlators whose effective mass is shown in the upper pane.

For our final results we pick fit ranges such that the determinations from different interpolators agree. For the determinations of the physical energy differences described in the next sections, results from correlated differences and ratio fits will be used as model variations in the AICc-based model averaging used to determine the central value and final combined uncertainties.

\subsection{A comparison to Lattice-NRQCD}

\begin{figure}[tb]
\centering
\includegraphics[scale=0.26]{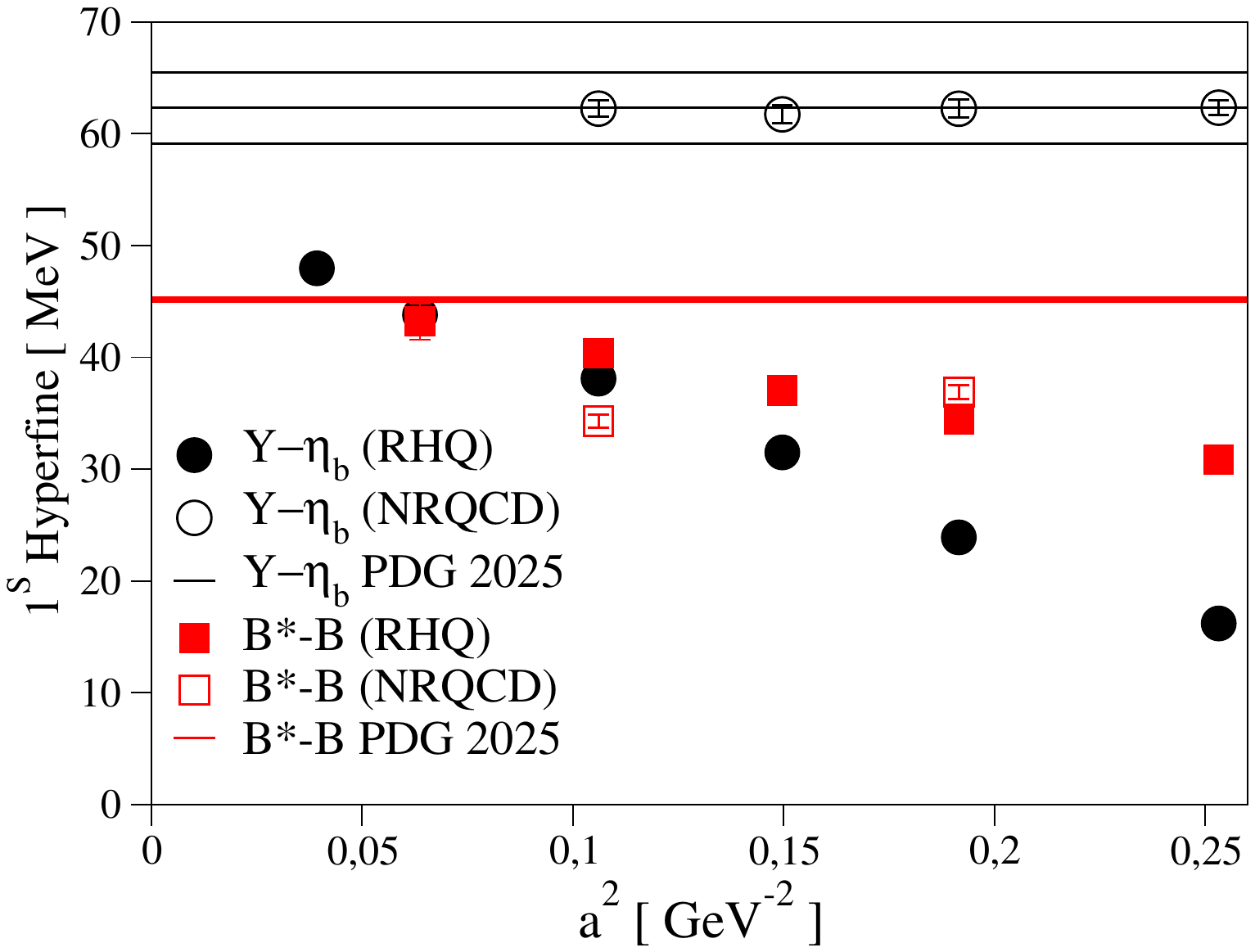}
\caption{A comparison of the hyperfine splittings we obtained for bottomonia and B-mesons using our RHQ action and our nonperturbatively tuned NRQCD action in \cite{Hudspith:2023loy}. Note that for the data from the RHQ action an extrapolation to the physical point will be performed in Sec.~\ref{sec:tests}.}\label{fig:hyperfine_NRQCD}
\end{figure}

We finish this section by comparing our hyperfine splittings at the $\text{SU}(3)_f$-symmetric point between our Lattice-NRQCD tuning of \cite{Hudspith:2023loy} and our RHQ determination presented here; this is shown in Fig.~\ref{fig:hyperfine_NRQCD}. As is evident, the discretisation effects for the $\Upsilon-\eta_b$ splitting are large for our RHQ action, and the current PDG value \cite{ParticleDataGroup:2024cfk} (62.3(3.2)) is completely reproduced for our Lattice-NRQCD action. Even though this splitting is part of our tuning in both cases, for the Lattice-NRQCD there is enough freedom in the parameters to make this agreement absolute. 

The large discrepancy from the continuum in the RHQ case is because the minimisation routine is  constrained by other parameters ($\kappa$ and keeping $c^2=1$) and can only limit the size of discretisation effects. It does appear that our bottomonia splitting will extrapolate to around the PDG result, albeit on the lower end. A simple $a^2$ and $a^4$ extrapolation of the 4 finest points gives 56.2(2.2), neglecting systematics from the absence of disconnected diagram contributions. There is some experimental tension in this quantity, with the most precise experimental determination \cite{Belle:2012fkf} (57.9(2.3)) being substantially lower than the others \cite{CLEO:2009nxu,BaBar:2009xir}. Our naive extrapolated result is consistent with the lattice determinations of \cite{Burch:2009az,Hatton:2021dvg,Cai:2026xja,RBC:2012pds}, all of which have central values below the PDG average.

For the B-mesons the situation is much better; the RHQ action introduces smaller discretisation effects and the splitting is much closer to the continuum value. For Lattice-NRQCD the situation is conversely much worse; the finer lattice spacing points to a smaller hyperfine splitting. This is clear evidence that one should never take a ``continuum limit" - i.e. an $a^2\rightarrow 0$-extrapolation of lattice NRQCD data (for our data this would give a hyperfine splitting of $31.1(1.5)$ MeV - almost 10$\sigma$ from the PDG \cite{ParticleDataGroup:2024cfk} fit $45.18(20)$ MeV). Even our strategy used in \cite{Hudspith:2023loy} of taking the midpoint of the two and quoting their difference as a systematic error leads to a value incompatible with experiment ($35.6(2.6)$ MeV or nearly 4$\sigma$ from the PDG value).
It is clear that using an RHQ action appears the more sensible thing to do for quantities that are sensitive to such $B$-meson splittings. This includes states such as the $T_{bb},\, T_{bbs}$ and $T_{bc}$ as well as potentially the exotic $B_s$-mesons considered in this work.

With our data it appears that the usage of Lattice-NRQCD should be constrained to states only containing heavy quarks, and for accurate reproduction of heavy-light continuum physics an RHQ prescription for the heavy quark should be preferred. At fixed lattice spacing it seems not possible to recreate both the heavy-onium and heavy-light spectrum simultaneously, only under a controlled continuum extrapolation. We will further illustrate our ability to reproduce continuum splittings from our heavy-quark action in the next section.
\section{Continuum tests for our action\label{sec:tests}}

\begin{table}[t]
\begin{tabular}{c|c|cc|cc}
\toprule
Name & $\beta$ & $\phi_\pi$ & $\phi_K$ & $m_\pi L$ & $m_K L$\\ 
\hline
A653      & 3.34 & 0.7803(70) & 0.7803(70) & 5.086 & 5.086 \\
A654      & 3.34 & 0.4860(77) & 0.9045(84) & 3.995 & 5.451 \\
GSI\_B650 & 3.34 & 0.2939(27) & 0.9603(56) & 4.106 & 7.421 \\
\hline
U103 & 3.40 & 0.7496(51) & 0.7496(51) & 4.330 & 4.330 \\
H101 & 3.40 & 0.7557(56) & 0.7557(56) & 5.829 & 5.829 \\
H102 & 3.40 & 0.5545(54) & 0.8421(63) & 4.926 & 6.126 \\
H105 & 3.40 & 0.3405(70) & 0.9462(63) & 3.884 & 6.474 \\
X150 & 3.40 & 0.3448(27) & 0.9441(27) & 4.873 & 8.062 \\
N101 & 3.40 & 0.3403(32) & 0.9365(30) & 5.832 & 9.676 \\
\hline
B450 & 3.46 & 0.7559(48) & 0.7559(48) & 5.140 & 5.140 \\
X451 & 3.46 & 0.3600(25) & 0.9351(21) & 4.419 & 7.121 \\
\hline
H200 & 3.55 & 0.7549(86) & 0.7549(86) & 4.331 & 4.331 \\
N202 & 3.55 & 0.7410(53) & 0.7410(53) & 6.444 & 6.444 \\
N203 & 3.55 & 0.5214(24) & 0.8539(27) & 5.402 & 6.913 \\
N200 & 3.55 & 0.3522(25) & 0.9382(31) & 4.432 & 7.234 \\
D200 & 3.55 & 0.1755(16) & 1.0126(29) & 4.164 & 10.003 \\
\hline
N300 & 3.70 & 0.7636(38) & 0.7636(38) & 5.073 & 5.073 \\
\botrule
\end{tabular}
\caption{$n_f=2+1$ CLS ensembles used in this work. Values of $\phi_\pi,m_\pi L,$ and $m_K L$ were determined in \cite{Ce:2022kxy} and \cite{Hudspith:2024kzk}, results for the ensemble X150 were determined as part of this project. The ensemble labeled \emph{GSI\_B650} has the CLS 2+1 flavor action but is not a CLS ensemble.}\label{tab:configs}
\end{table}

As we performed our heavy-quark tuning only on bottomonia, the splittings of $B$- and $B_s$-mesons are predictions of our calculations. Some of these splittings are well-known experimentally and provide an important benchmark for the validity of our continuum limit in heavy-light systems with the heavy valence quark from our RHQ parameters.

For this section and the evaluation of the exotic $B_s$-mesons we will use the $n_f=2+1$ CLS configurations with constant trace of the quark-mass matrix, as listed in Tab.~\ref{tab:configs}. More details on these ensembles can be found in \cite{Bruno:2014jqa,RQCD:2022xux}. We span pion masses from the $\text{SU}(3)_f$-point, around $420$ MeV, to $200$ MeV (D200). The ensembles used in this work span a large range of $m_\pi L$ and $m_K L$ for adequate control over finite-volume effects. For extrapolations to the physical pion mass we use the dimensionless analog $\phi_\pi = 8t_0 m_\pi^2$, comparing it to the pion mass from the ``Edinburgh Consensus" and the $n_f=2+1$ value of $t_0$ from FLAG \cite{FlavourLatticeAveragingGroupFLAG:2024oxs}. 

We describe the 1S hyperfine splittings and the exotics by the following generic fit Ansatz (defining $\delta\phi_\pi = \phi_\pi^\text{latt}-\phi_\pi^\text{phys}$- the difference between our lattice-measured $\phi_\pi$ and the continuum limit value formed from the FLAG results),
\begin{widetext}
\begin{equation}\label{eq:master_fitform}
\Delta(a^2,\delta\phi_\pi,m_\pi L, m_K L) = 
\Delta(0,0,\infty,\infty)\left( 1 + Aa^2 + B\delta \phi_\pi + Ce^{-m_\pi L} + D e^{-m_K L}\right),
\end{equation}
\end{widetext}
with free fit parameters $\Delta(0,0,\infty,\infty),A,B,C,$ and $D$. Typically, overfitting our data is observed when too many parameters are used and we use a weighted average by the AICc \cite{burnham2002model} with an additional penalty parameter $N$ \cite{Borsanyi:2020mff} (to incorporate fits over reduced data sets) or variations e.g. in \cite{Jay:2020jkz, Ce:2022kxy}, to penalize fits with many parameters. Throughout this work we will refer to this procedure \textit{as} the AICc for brevity. When forming the AICc-weighted average we use the full bootstrapped fit parameter $\Delta(0,0,\infty,\infty)$ to propagate the error as it is very correlated between various fits and a typical naive error propagation formula would not be suitable. We apply the additional constraint that for the $B^*-B$ splitting $D=0$ as a kaon-based finite-volume term in a quantity that only has valence light $u$ and $d$ quarks seems spurious and should be strongly suppressed.

\begin{figure}[tb]
\centering
\includegraphics[scale=0.45]{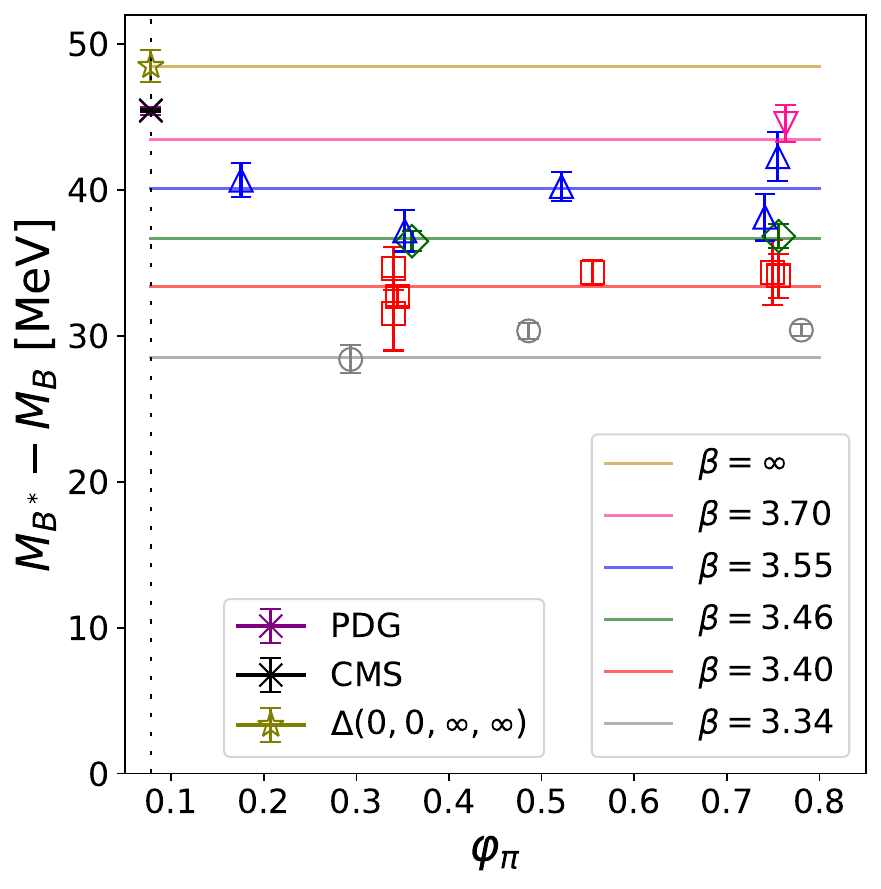}
\caption{Chiral-continuum extrapolation of the hyperfine splitting in the $B$ system. We display the results for the model with the highest weight in the AICc, as described in the text (Model A from correlated differences with a cut in lattice spacing). Horizontal lines indicate the fit result for the given $\beta$, with parameters $\Delta=48.5(1.1)$ and $A=-1.6(0.1)$, which has $\chi^2/\mathrm{dof}=1.1$. Ensembles with $\beta=3.34$ are disregarded due to this particular lattice-spacing cut, and appear as grey symbols.}
\label{fig:hyperfine_B}
\end{figure}

We incorporate the errors on $a, \phi_\pi, m_\pi,$ and $m_K$ by creating random Gaussian bootstrap resamples with width corresponding to their measured variances. We note that there is very little correlation between $m_\pi$ and the B-meson mass on ensembles where we have data for both, and treating them as uncorrelated is suitable. We do not have our own measured pion masses for all ensembles, so we use literature values for consistency. In addition to analyzing the full data set, we apply data cuts that remove the coarsest boxes A653, A654 and $\mathrm{GSI\_B650}$ ($\beta > 3.34$), the SU(3)-symmetric boxes A653, U103, H101, B450, H200, N202 and N300 ($m_{\pi} < 400$ MeV) or boxes U103 and H200, which have small physical volumes ($L > 2.2$ fm).

\begin{figure}[tb]
    \centering
    \includegraphics[width=0.4\textwidth]{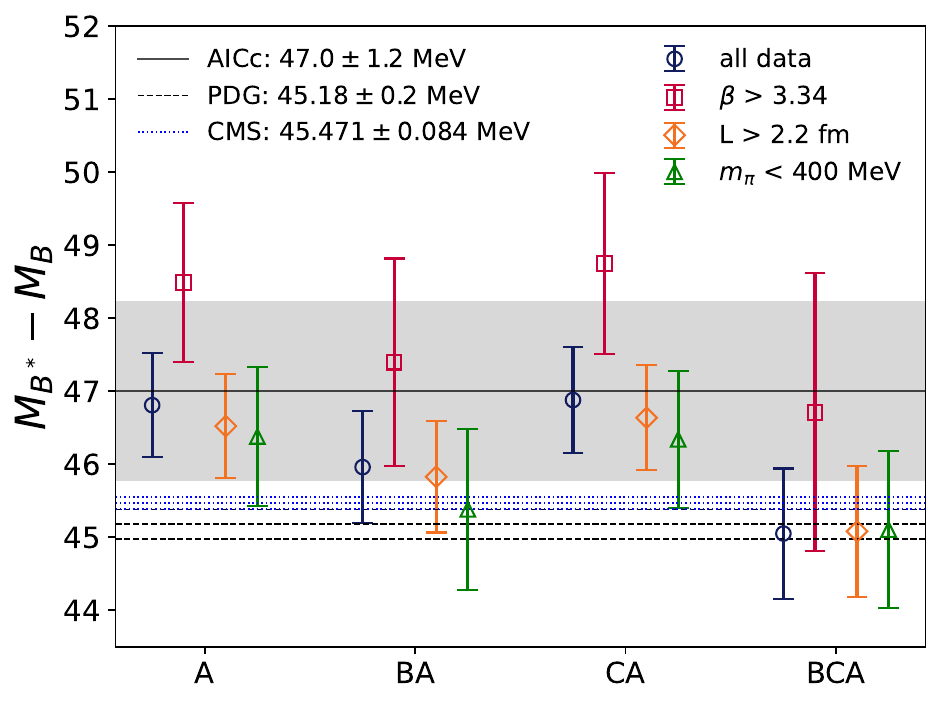}\\
    \includegraphics[width=0.4\textwidth]{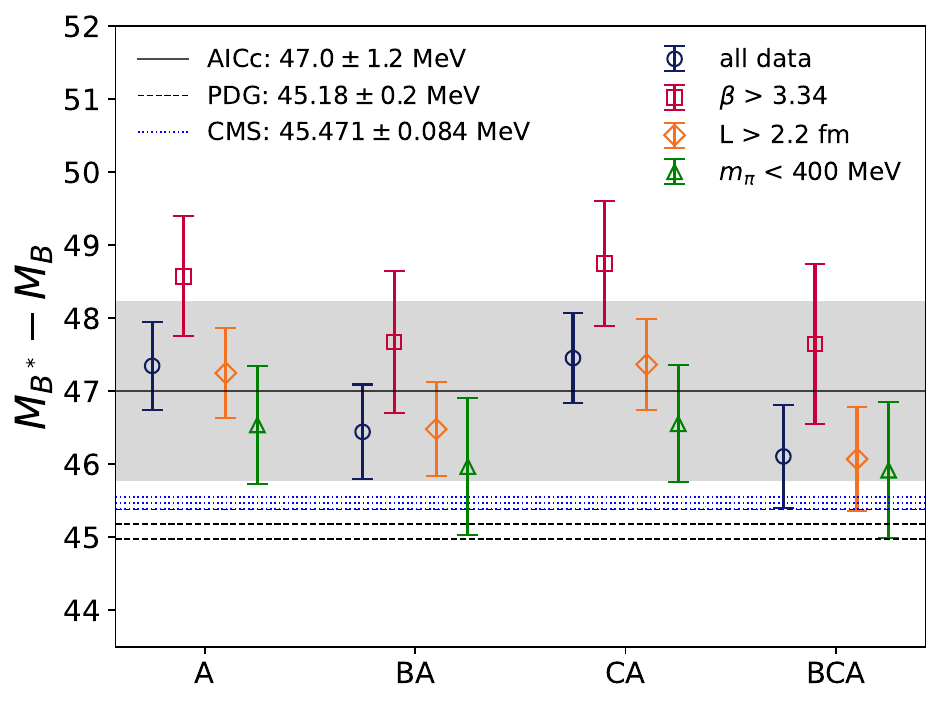}
\caption{Fit-model variation of the hyperfine splitting in the $B$ system. The x-axis label shows the active parameters in the chiral--continuum--infinite-volume fits (from Eq.~\ref{eq:master_fitform}). For each fit type the results are shown including those with additional cuts to the data (shifted for clarity). The grey band indicates the final AICc uncertainty of our estimate, while the dashed and dotted lines show the experiment results from the PDG \cite{ParticleDataGroup:2024cfk} and from CMS \cite{CMS:2025byz}. The top pane shows results from correlated differences, while the bottom shows results from ratio fits.}
\label{fig:hyperfine_B_fitvar}
\end{figure}

\subsection{The \texorpdfstring{$B$}{B}-meson 1S hyperfine splitting}

As a first quantity of interest, we determine the hyperfine splitting in the $B$-meson system $M_{B^\star}-M_B$. For this, we use standard interpolating fields for the vector and pseudoscalar states, although we find those with $\gamma_t\gamma_5$ and $\gamma_i\gamma_t$ have less excited state contamination than those without the $\gamma_t$, as can be seen in Fig.~\ref{fig:hyperfine_B_Meff_Stab}. The energies extracted in MeV - both from correlated differences and from the ratios of correlation functions - can be found in Table \ref{tab:all_results}. Employing the extrapolation formula in Eq. \ref{eq:master_fitform}, we depict the model with the highest weight in Figure \ref{fig:hyperfine_B} as an example. Here, the pion mass dependence is flat, finite-volume terms are deactivated, and the only correction parameter considered is the $\mathcal{O}(a^2)$ one- A.

\begin{figure}[tb]
\centering
\includegraphics[scale=0.45]{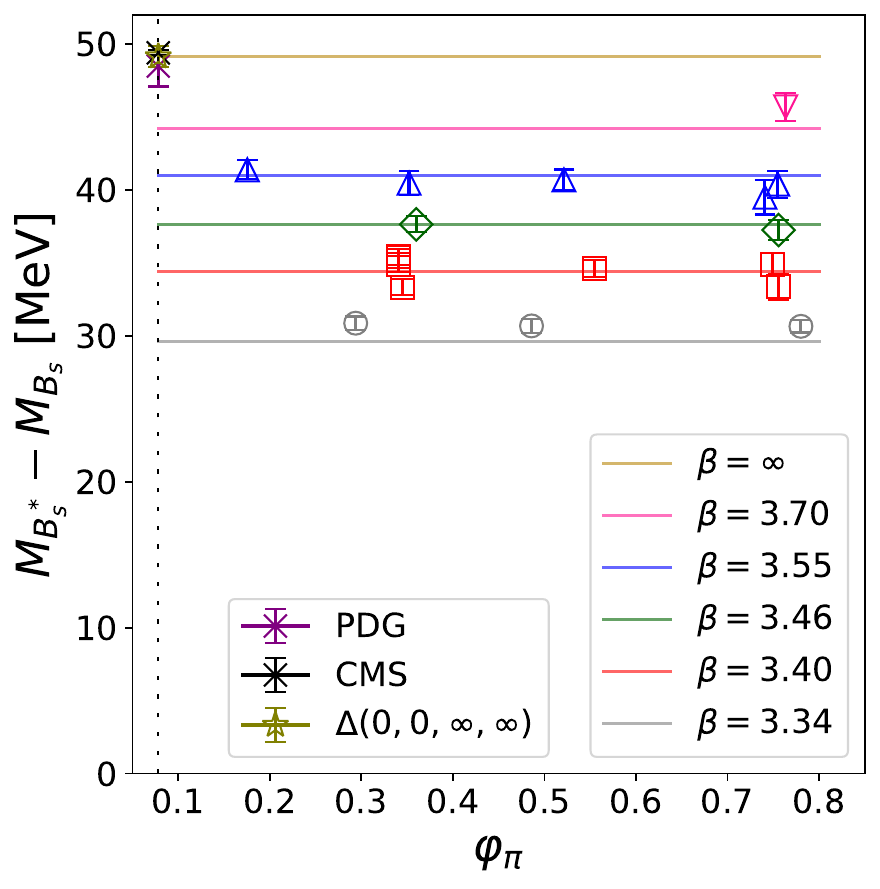} 
\caption{Chiral-continuum extrapolation of the hyperfine splitting in the $B_s$ system. We display the results for the model with the highest weight in the AICc, as decribed in the text (this time Model A from the ratio mathod with a cut in lattice spacing). Horizontal lines indicate the resulting fit for the given $\beta$, with parameters $\Delta=49.2(0.7)$, $A=-1.6(0.1)$ and $\chi^2/\mathrm{dof}=1.2$. Ensembles with $\beta=3.34$ are disregarded due to this particular lattice-spacing cut, and appear as grey symbols.}
\label{fig:hyperfine_Bs}
\end{figure}

The results of our various fits and data cuts are shown in Fig. \ref{fig:hyperfine_B_fitvar} and enter our AICc-determination of the full uncertainty of the physical splittings in the combined chiral-continuum and infinite-volume limit. It appears that the simplest fit with two parameters $\Delta(0,0,\infty,\infty)$ and A is preferred and removing the coarsest lattice-spacing data pulls our model average away from the PDG \cite{ParticleDataGroup:2024cfk} average and the very precise CMS determination \cite{CMS:2025byz}. This results in our final weighted-average result,
\begin{equation}
M_{B^\star}-M_B = 47.0(1.2)(0.3)_\text{Iso}(0.4)_\text{Tuning} \text{ MeV}.
\end{equation}
This value is shown as a grey uncertainty band in Fig. \ref{fig:hyperfine_B_fitvar}. The first error is from our AICc averaging procedure -- a combination of statistical error and systematics from fit variations. Our resulting uncertainty is much larger than that of experiment and the central value is in reasonable ($1\sigma)$) but not perfect agreement. Our ``Tuning" systematic is estimated from the average model uncertainty in $c_B$, which is likely the dominant parameter for the hyperfine splitting as is generically expected from Lattice-NRQCD. To determine this we average the percentage uncertainty for that parameter from Tab.~\ref{tab:bparams_v2} and apply that to our final central result. For the ``Iso" (Isospin and QED) systematic we take the full difference of the $B^+-B^0$.

\subsection{The \texorpdfstring{$B_s$}{Bs} 1S hyperfine splitting}

\begin{figure}[tb]
    \centering
    \includegraphics[width=0.4\textwidth]{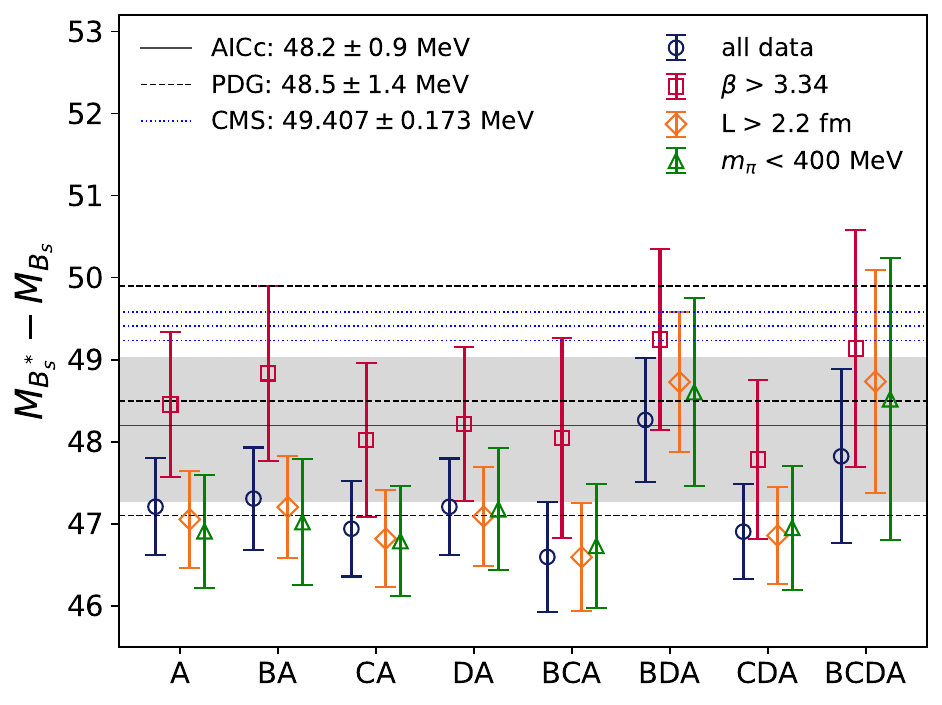}
    \includegraphics[width=0.4\textwidth]{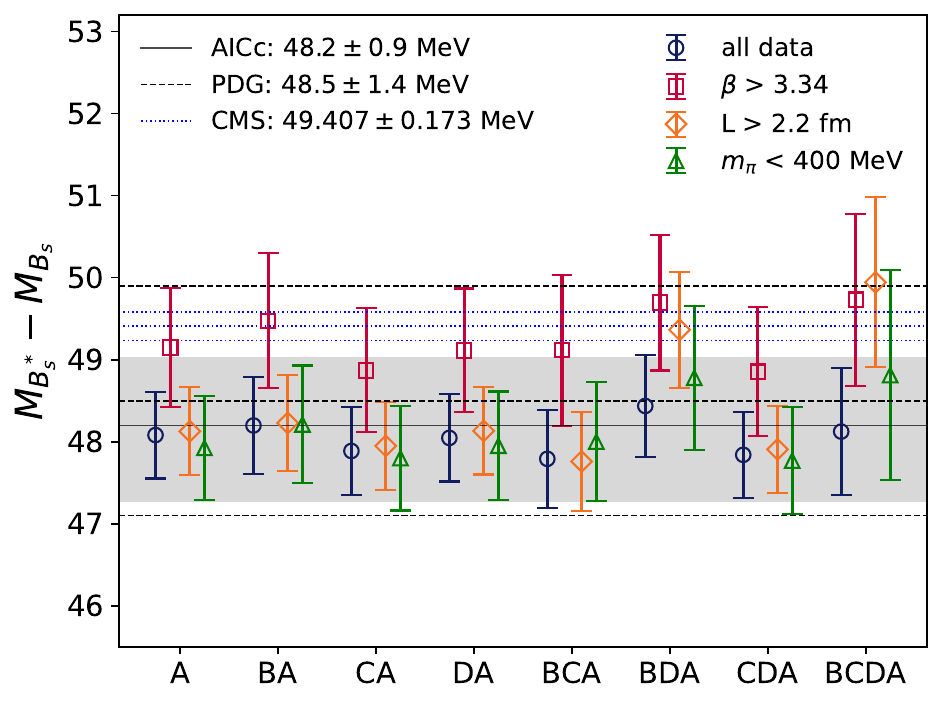}
\caption{Fit-model variation of the hyperfine splitting in the $B_s$ system. The x-axis label shows the active parameters in the chiral--continuum--infinite-volume fits (from Eq.~\ref{eq:master_fitform}). For each fit type the results are shown including those with additional cuts to the data (shifted for clarity). They grey band indicated the final AICc uncertainty of our estimate, while the dashed and dotted lines show the experiment results from the PDG \cite{ParticleDataGroup:2024cfk} and from CMS \cite{CMS:2025byz}. The top pane shows results from correlated differences, while the bottom shows results from ratio fits.}
\label{fig:hyperfine_Bs_fitvar}
\end{figure}

The $B_s^*-B_s$ analysis follows that of the preceding section. Again we find that the currents with a $\gamma_i\gamma_t$ have slightly less excited state contamination, and from monitoring the $\gamma_i$ and $\gamma_i\gamma_t$ corelators we can identify a reasonable plateau where the two agree. Away from the $\text{SU}(3)_f$-point the $B_s$ and $B_s^*$ are considerably more statistically-precise in comparison to the $B$ and $B^*$, respectively. However, their excited-state contamination is worse and they typically take longer to plateau in their effective mass.

As we expect pion and kaon loops to contribute to our 1S hyperfine splitting for the $B_s$ we include both pion- and kaon-based finite-volume terms for our fit in Eq.~\ref{eq:master_fitform} in the AICc analysis over possible fits and data cuts. We find little difference in our final determination from either of these finite-volume terms $C$ or $D$. In fact, a very simple fit form with a single parameter $\propto a^2$ is preferred (see Fig.~\ref{fig:hyperfine_Bs}). We see little indication of $m_\pi^2$ behaviour let-alone higher-order pion mass corrections such as $m_\pi^3$.

\clearpage

From Fig.~\ref{fig:hyperfine_Bs_fitvar} we again observe reasonable consistency between our determination of the $B_s^*-B_s$ hyperfine splitting and the average of the PDG and that of CMS. There is more consistency between our \emph{ratio fits} analysis under data cuts, than our \emph{correlated difference} analysis, in particular under the removal of our coarsest boxes. Our weighted-average result is
\begin{equation}
M_{B^\star_s}-M_{B_s} = 48.2(0.9)(0.3)_\text{Iso}(0.4)_\text{Tuning} \text{ MeV}.
\end{equation}
The estimated tuning uncertainty is the same as that for the $B$-meson hyperfine.

\subsection{The \texorpdfstring{$B_s-B$}{Bs-B} splitting}

\begin{figure}[tb]
\includegraphics[scale=0.26]{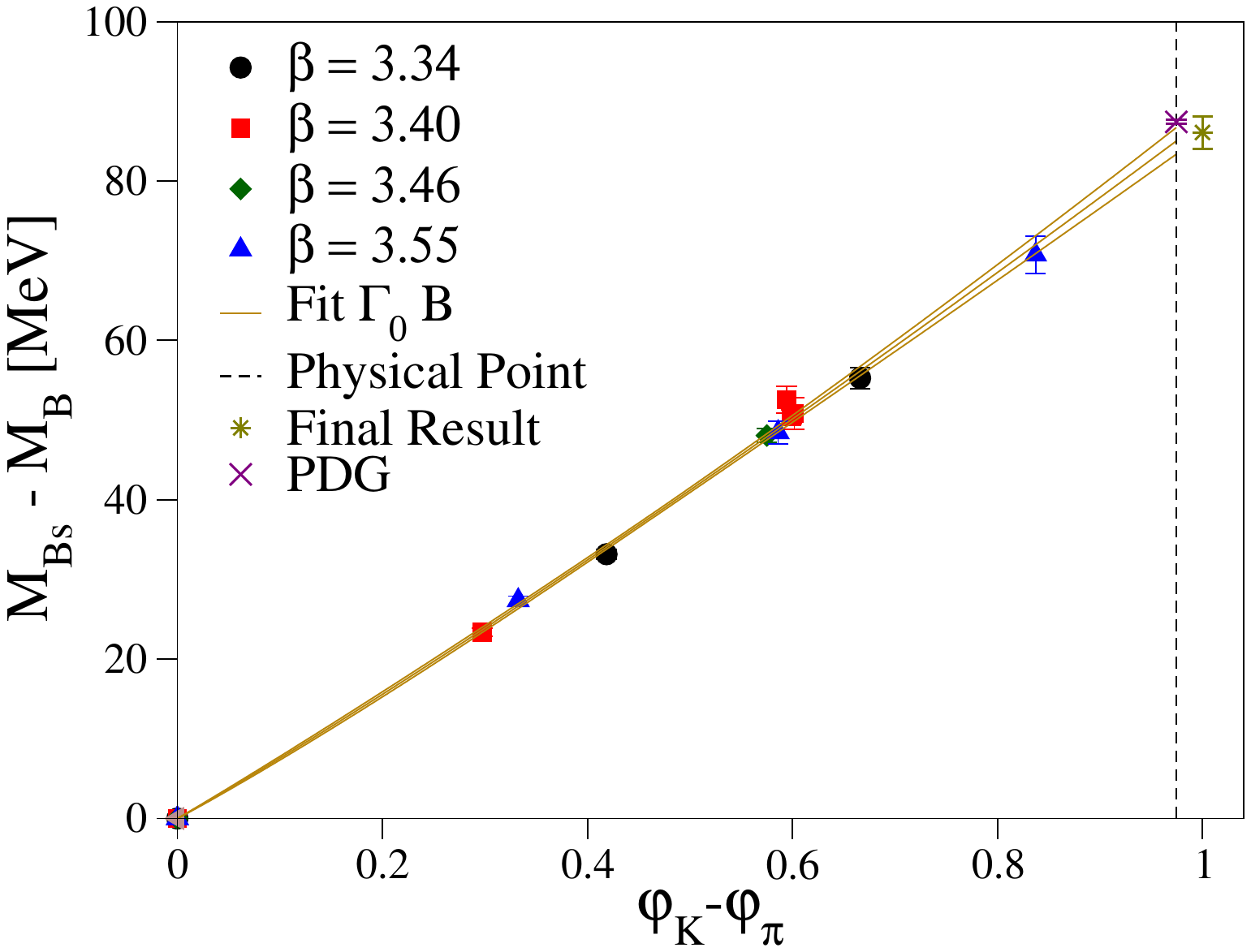}
\caption{Our best-weighted fit for the splitting $M_{B_s}-M_B$ and the final weighted-average result (offset for clarity) in comparison to the PDG average.}\label{fig:BsmB}
\end{figure}

For the mass splitting between the pseudoscalar $B_s$ and $B$ mesons we have additional constraints that make the fit-form of Eq.~\ref{eq:master_fitform} inadequate. Naturally, one might expect that this splitting to be proportional to the strange-light mass splitting $m_s-m_\ell$ and to leading order this is the case. Inevitably, this splitting has to vanish when $m_s=m_\ell$, and this is what we bake into our fit. Noticing that the quantity $\phi_K-\phi_\pi \propto m_s-m_\ell$ this is a natural x-axis, we thus parameterize our $B_s-B$ splitting data with the Ansatz:
\begin{widetext}
\begin{equation}
\Gamma(a^2,\phi_K-\phi_\pi,m_\pi L, m_K L) = (\phi_K-\phi_\pi)\left( \Gamma_0 + A a^2 + B\sqrt{\phi_K-\phi_\pi} + C e^{-m_\pi L} + D(\phi_K-\phi_\pi)\right).
\end{equation}
\end{widetext}

This guarantees that the fit and our data go through the origin, and we need only evaluate this fit function at the FLAG-determined physical point for $\phi_K-\phi_\pi$. 
As the $\text{SU}(3)_f$ point provides the largest handle on $m_K L$ and those data are all exact zeros in our description, any $m_KL$ effects are therefore suppressed and the finite-volume effects are likely dominated by the the pion. There is clear curvature in our data and we model this with either a cubic (in $m_K$) $B$ or quartic $D$ correction. The former motivated by Heavy Meson Chiral Perturbation Theory \cite{Goity:1992tp} and the latter term expected as a higher-order analytic mass correction. Our fit only slightly favors curvature $\propto m_K^3$, as opposed to $m_K^4$. 

Our highest-weighted fits always contain either $B$ or $D$, but their combination has a dramatically suppressed weight. Overall the two highest-weighted fits from the AICc have parameters $\Gamma_0,B$ and $\Gamma_0,D$, and the next-highest include $\Gamma_0$, $B$ or $D$, and the finite-volume term $C$. Our final weighted-average result is:
\begin{equation}
M_{B_s} - M_B = 86.1(2.0)(0.3)_\text{Iso} \text{ MeV},
\end{equation}
with the first error being statistical and the second our expected systematic from Isospin breaking, which we estimate as the full difference between the $B^\pm$ and $B^0$ masses. While our determination is completely consistent with the current PDG average of $87.42(24)\text{ MeV}$ \cite{ParticleDataGroup:2024cfk}, we would like to stress that the systematic uncertainty from mistuning of the CLS trajectory with constant trace of the quark-mass matrix might be relevant for this observable. Quantifying this uncertainty or correcting it would require adding ensembles from the other CLS quark-mass trajectories \cite{RQCD:2022xux}, which is beyond the scope of the current project. Figure~\ref{fig:BsmB} shows the fit with two parameters, $\Gamma_0$ and $B$, as this has the largest weight, as well as our final value including the quantified systematics.

\begin{figure}[b]
\includegraphics[scale=0.26]{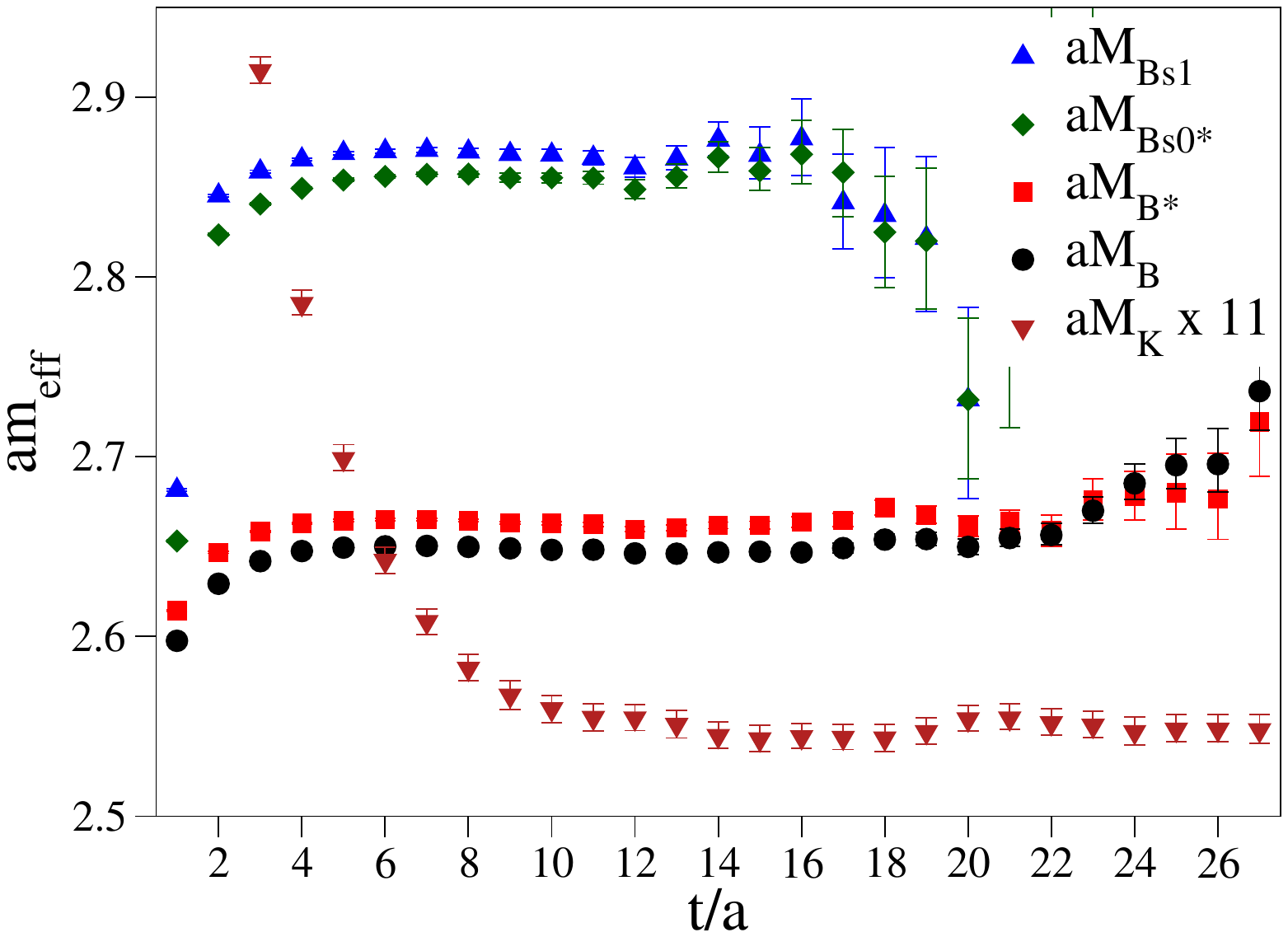}
\caption{Effective masses on ensemble \emph{GSI\_B650} for the $B_{s0}^*$, the $B_{s1}$, the $B$ from the $\gamma_5,\gamma_5$ correlator and the $B^*$ from the $\gamma_i,\gamma_i$ correlator, as well as the kaon multiplied by a factor of 11 to fit nicely on the figure.}\label{fig:effmassB650}
\end{figure}

\begin{figure*}[tb]
\includegraphics[width=0.32\textwidth]{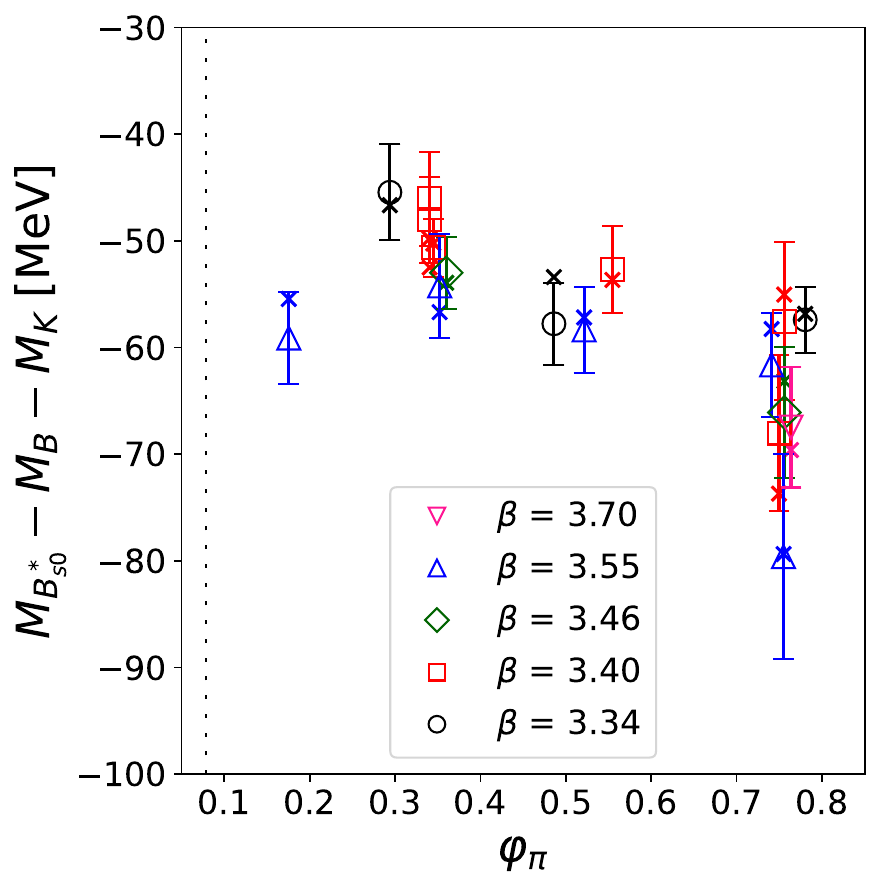}
\includegraphics[width=0.32\textwidth]{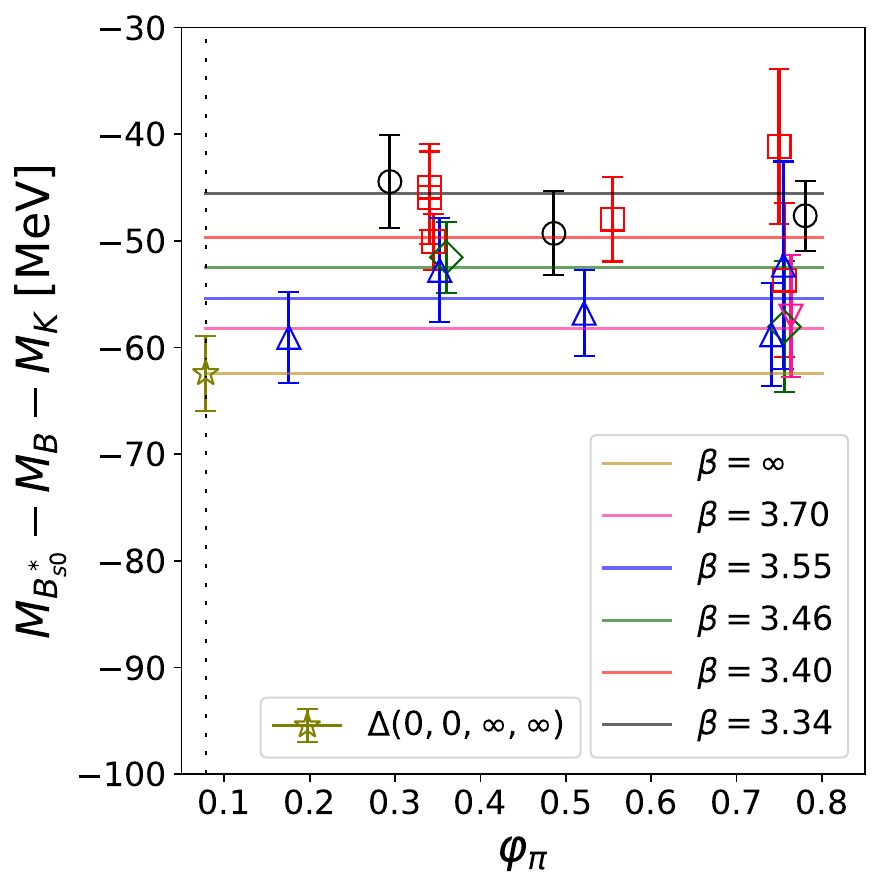}
\includegraphics[width=0.32\textwidth]{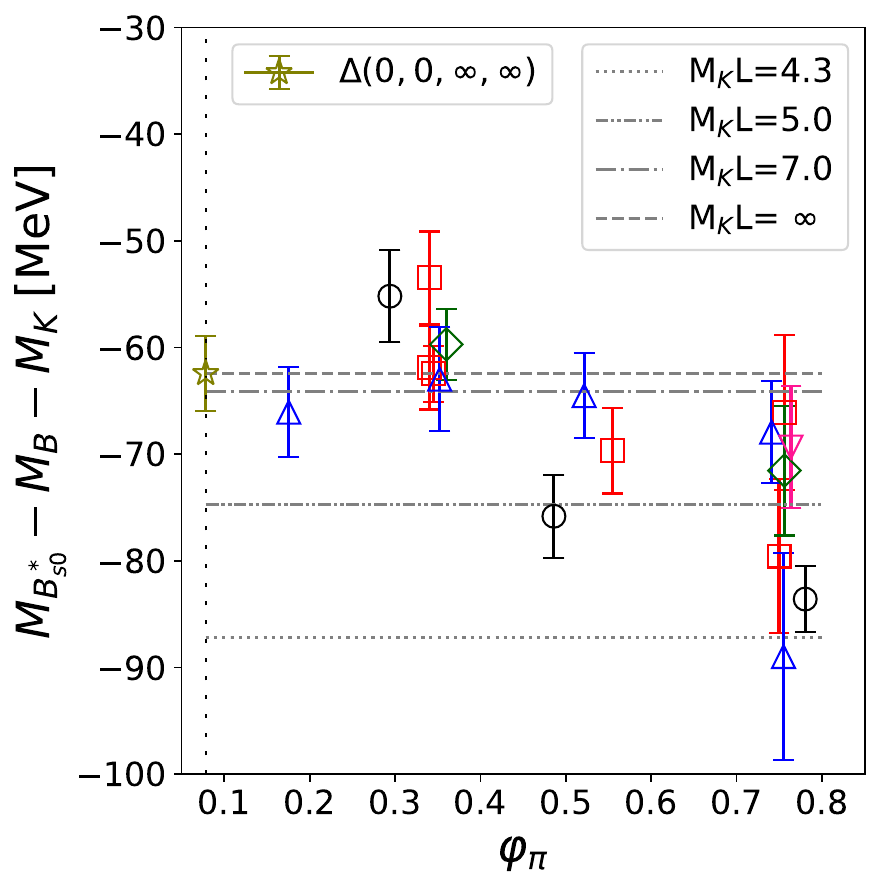}
\caption{Binding energy for the $B_{s0}^*$ meson with respect to the $BK$ threshold. The left pane shows the $B_{s0}^*$ binding energy measured on the CLS ensembles in Table \ref{tab:configs}. The middle pane shows the resulting continuum extrapolation for the fit model with the highest AICc weight (parameters $\Delta(0,0,\infty,\infty)= -62.4(3.5)$, $D = 29.2(6.1)$, $A = -1.1(0.3)$, $\chi^2/dof = 0.4$) after subtracting the finite-volume effects. The right pane shows the finite-volume effects after subtracting the discretisation term.}
\label{fig:Bs0_extrapolation}
\end{figure*}

\section{Exotic \texorpdfstring{$B_s$}{Bs} meson masses}

We have shown that our action can reproduce the experimental values for heavy-light hyperfine splittings to a reasonable level of accuracy. This provides ample evidence that continuum predictions of hitherto experimentally-unknown quantities from our action are reliable. This is where we will turn in this section, to predictions of the scalar $B_{s0}^*$ and axial-vector $B_{s1}$ mesons. We will use the simple local meson operators for both,
\begin{equation}\label{eq:exotic_ops}
O_{B_{s0}^*} = \bar{b}s, \qquad O_{B_{s1},i} = \bar{b}\gamma_5\gamma_i s.
\end{equation}
After forming the correlator and measuring the associated mass in lattice units we subtract the relevant lowest-lying two meson thresholds at rest $BK$ and $B^*K$ respectively. An investigation into possible close-by levels for the $B_{s1}$ and the reliability of the obtained binding energy is presented in App.~\ref{app:gevp}. For the model average of these threshold-subtracted exotic states we introduce an additional fit parameter $Ea^3$ to Eq.~\ref{eq:master_fitform} in order to incorporate potentially higher-order discretisation effects along with our typical $\mathcal{O}(a^2)$ term. In fits with this extra parameter we deem it sensible to omit the cut in $\beta$ when forming the model average, as our coarsest ensembles should be the most sensitive to this term.

Fig.~\ref{fig:effmassB650} illustrates the quality of our data on the ensemble \emph{GSI\_B650}. It is clear that the $B_{s0}^*$ and $B_{s1}$ are quite similar in their masses, plateau earliest, and are most effected by statistical noise at reasonably short times. The $B$ and $B^*$ have a zenith where the mass turns over; their effective masses plateau slowly and fairly late, indicating nearby alternating-signed amplitudes in the tower of excited states. Finally, the kaon takes the longest to plateau. This behaviour is universal throughout our ensembles and would make a \emph{ratio fits} determination of the mass splitting $\Delta$ unsuitable due to the late saturation of the kaon, $B$, and $B^*$ in relation to the $B_{s0}^*$ and $B_{s1}$. This could lead to ``fake plateaus" and therefore we stick to performing a \emph{correlated differences} analysis for the $B_{s0}^*$ and $B_{s1}$.

\begin{figure}[tb]
    \centering
    \includegraphics[width=0.95\columnwidth]{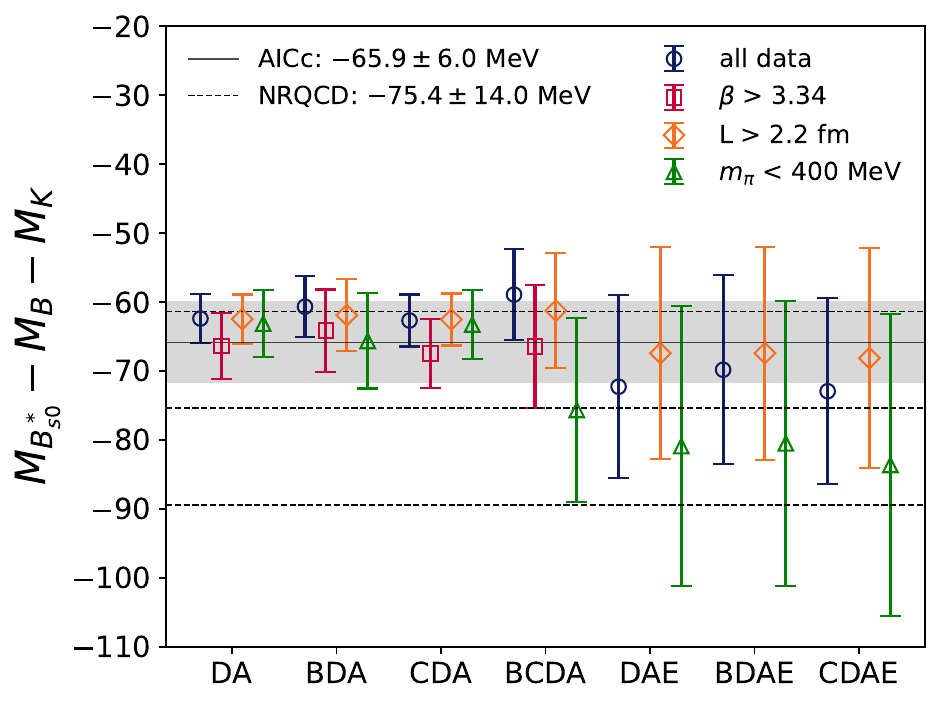}
\caption{The fit variations tried and the corresponding model average for the binding energy of the $B_{s0}^*$.}
\label{fig:Bs0_fitvar}
\end{figure}

\subsection{The \texorpdfstring{$j=\frac{1}{2}$ $B_{s0}^*$}{j=1/2 Bs0*}}

Fig.~\ref{fig:Bs0_extrapolation} shows a triptych: the leftmost panel illustrates our determined binding energy (with stars illustrating the central value prediction of the best fit we obtain from the AICc), the central panel shows the continuum extrapolation after finite-volume effects of the form $e^{-m_KL}$ have been subtracted, and the right panel shows the data after the discretisation effect term $\mathcal{O}(a^2)$ has been subtracted. It is clear that the dominant effect in our data is that of finite-volume effects from the kaon, in accordance with our previous Lattice-NRQCD study \cite{Hudspith:2023loy}. 

\begin{figure*}[t]
\includegraphics[width=0.30\textwidth]{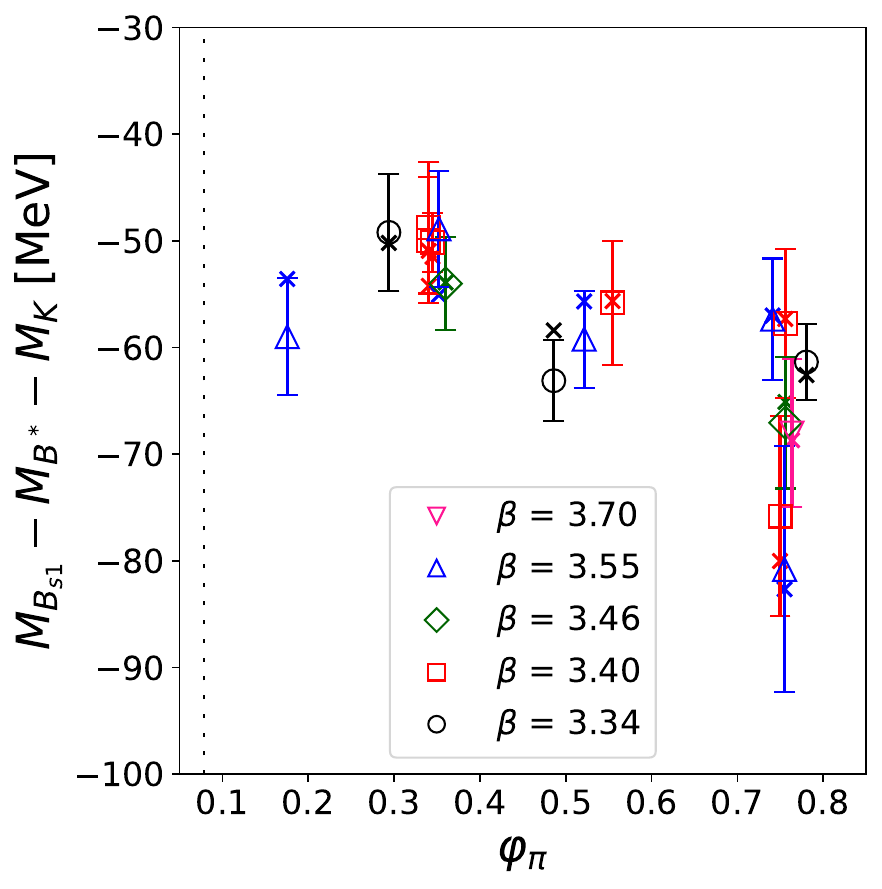}
\includegraphics[width=0.30\textwidth]{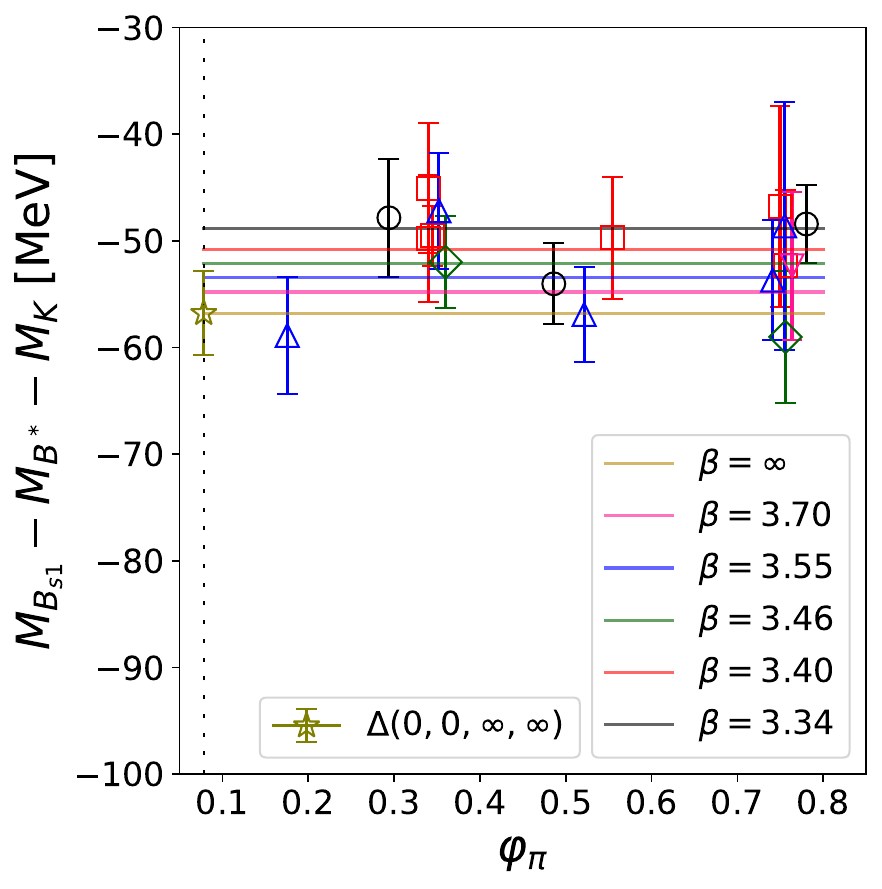}
\includegraphics[width=0.30\textwidth]{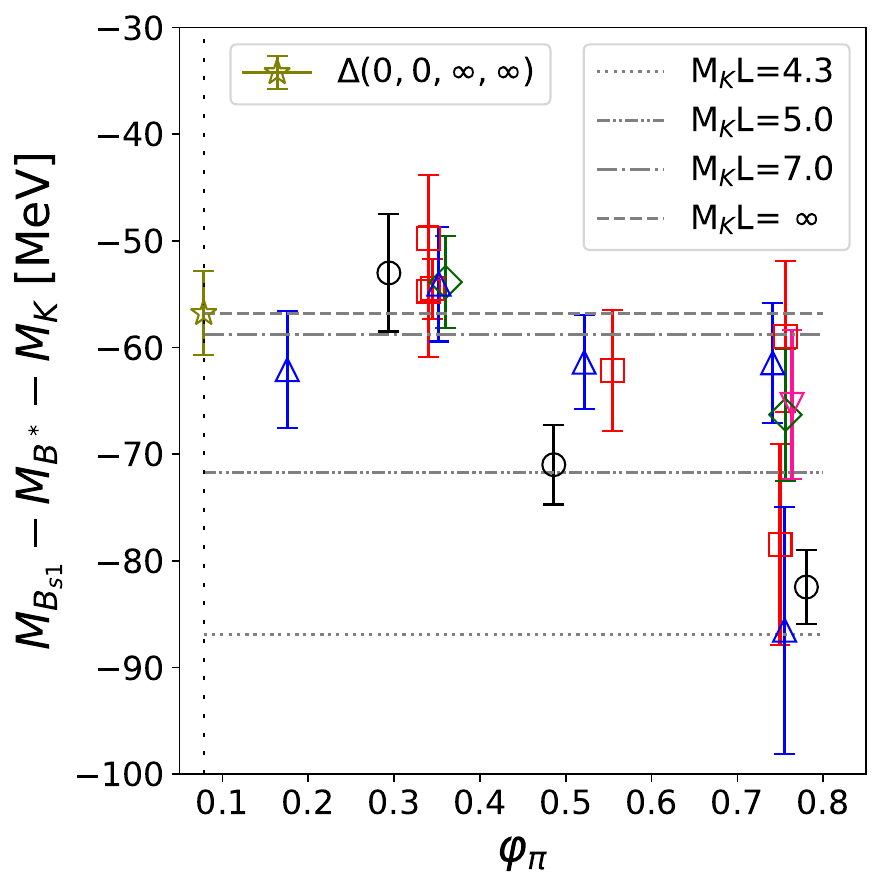}
\caption{Binding energy for the $B_{s1}$ meson with respect to the $B^*K$ threshold. The left pane shows the $B_{s1}$ binding energy measured on the CLS ensembles in Table \ref{tab:configs}. The middle pane shows the resulting continuum extrapolation for the fit model with the highest AICc weight (parameters: $\Delta(0,0,\infty,\infty) = -56.8(3.9)$, $D = 39.1(0.2)$, $A = -0.6(0.3)$, $\chi^2/dof = 0.4$) after subtracting the finite-volume effects. The right pane shows the finite-volume effects after subtracting the discretisation term.}
\label{fig:Bs1_extrapolation}
\end{figure*}

We are particularly sensitive to $e^{-m_K L}$ effects from the boxes at the flavor-$\text{SU}(3)$ point, U103 and H200, which have $m_\pi L = m_K L \approx 4.3$. We note that a fit to $e^{-m_\eta L}$, which could be expected from \text{SU(3)} $\chi_\text{PT}$, with $m_\eta$ from the GMOR relation is not favored. There is a clear clustering of data points away from the heavy pion-mass data, which span a wide range of $m_\pi L$ indicating that pion-based finite-volume effects of the form $e^{-m_\pi L}$ are negligible. The term we use to fit discretisation effects is small in comparison to the finite volume term and has a negative slope whereas the finite volume effects have a positive parameter, leading to some cancellation of the two effects amongst our points and a deeper binding in continuum. 

Figure~\ref{fig:Bs0_fitvar} shows the model average over the various fit prescriptions and cuts performed. We also show the value, including the full statistical and systematic uncertainty of our previous Lattice-NRQCD calculation \cite{Hudspith:2023loy}.
Unlike in the cases for the hyperfine splittings there is little pull from any of the cut variations and the simplest fit (that of $DA$) describes the data very well. A curious effect happens for the fit $CDA$ in comparison to $BCDA$ where the statistical resolution on $\Delta$ is reduced, presumably due to the $e^{-m_\pi L}$ term roughly approximating the underlying small slope in pion mass that is better described by the parameter $B$. Introducing an $\mathcal{O}(a^3)$ correction term is somewhat disfavored by the model average and the resulting error is enhanced suggesting higher-order discretisation effects are not evident at our level of precision.

\subsection{The \texorpdfstring{$j=\frac{1}{2}$ $B_{s1}$}{j=1/2 Bs1}}

The axial-vector $B_{s1}$ fairs similarly to the scalar; the dominant feature in our data is that of kaon-based finite-volume corrections. The fit parameter describing discretisation effects $A$ is about half that of the $B_{s0}^*$. Here, Fig.~\ref{fig:Bs1_extrapolation} gives the $B_{s1}$ analog to Fig.~\ref{fig:Bs0_extrapolation}, and the two behave very similarly.

\begin{figure}[tb]
    \centering
    \includegraphics[width=0.95\columnwidth]{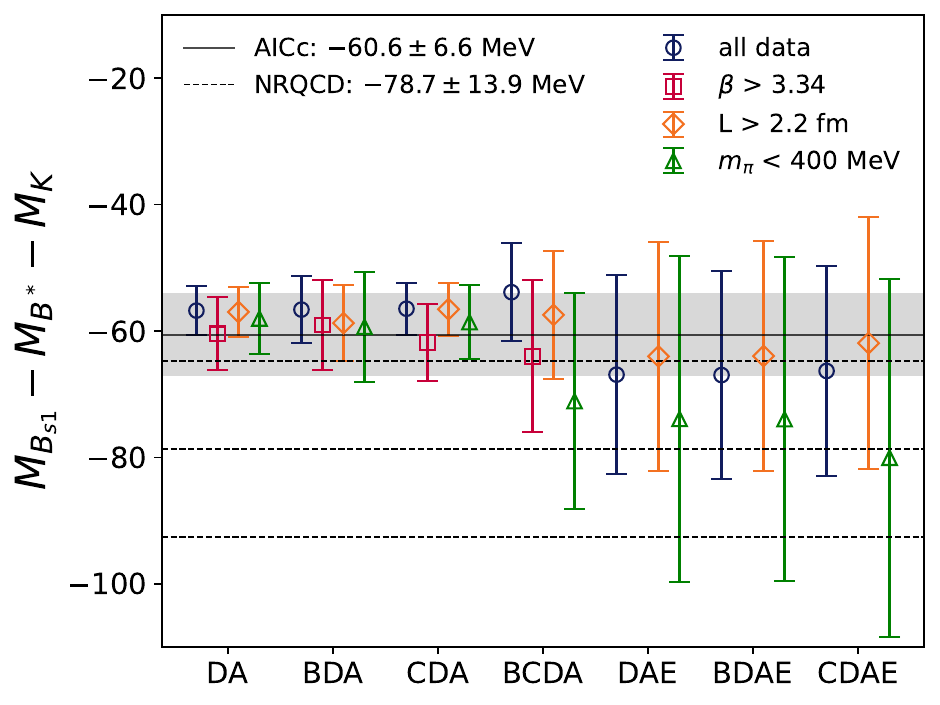}
\caption{The fit variations tried and the corresponding model average for the binding energy of the $B_{s1}$.}
\label{fig:hyperfine_Bs1_fitvar}
\end{figure}

Figure~\ref{fig:hyperfine_Bs1_fitvar} shows our final model-average result for the binding energy of the $B_{s1}$ state, including all the variations of fit models and cuts. Much like in the $B_{s0}^*$ our final result lies at the upper end of our previous lattice-NRQCD calculation, indicating a somewhat shallower binding than predicted there. We see good consistency between all of these variations and the simplest fit-form $DA$ is preferred.
\section{Conclusions}

\begin{figure*}[tb]
    \begin{center}
    \includegraphics[width=0.65\textwidth]{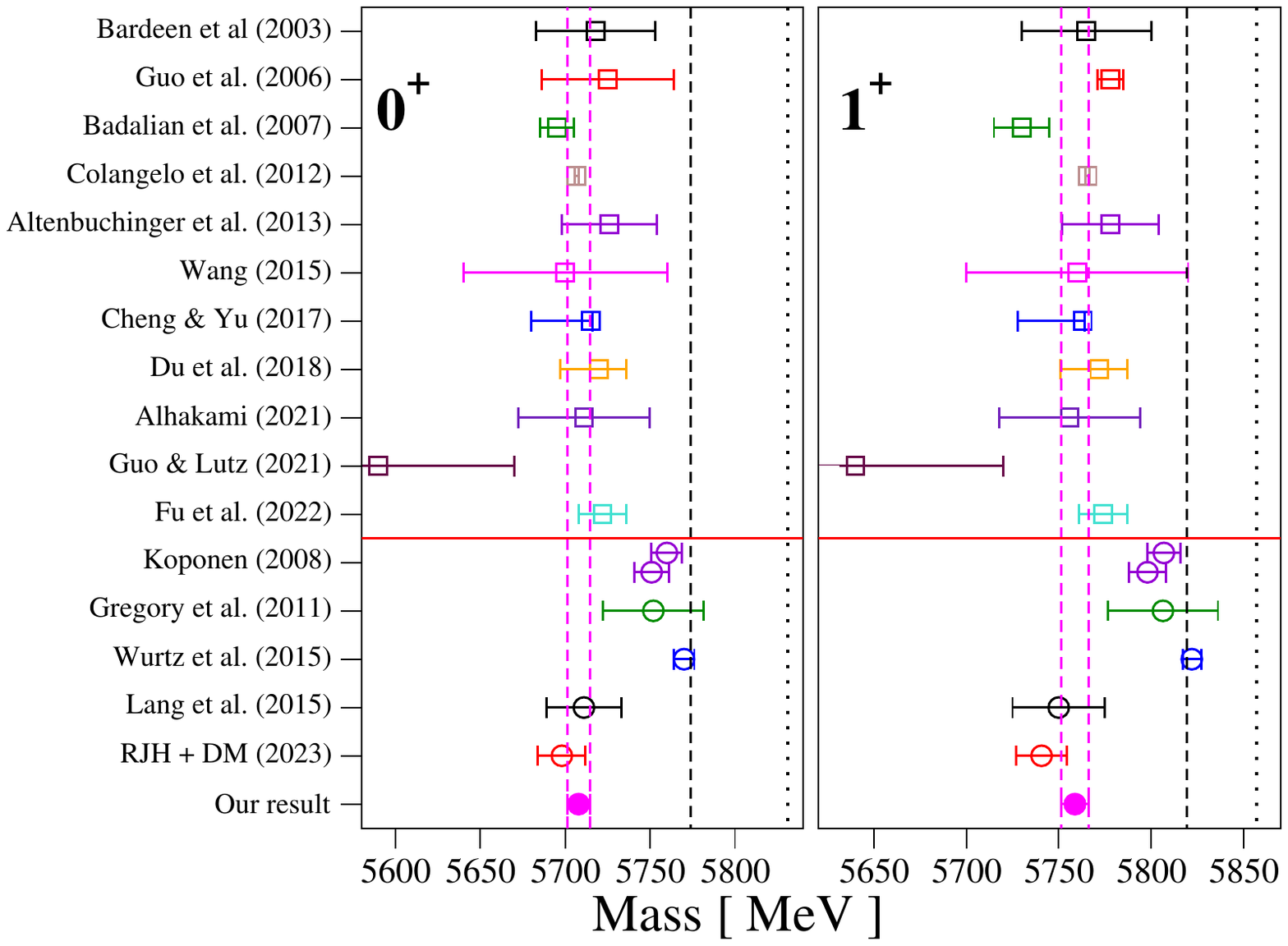}
    \end{center}
    \caption{Various predictions of the $B_{s0}^*$ (left pane) and $B_{s1}$ (right pane) masses from models or EFT+model calculations (rows above the red lines) and Lattice QCD calculations (rows below the red lines). The vertical dashed black lines indicate the $BK$ and $B^\star K$ thresholds respectively and the dotted lines are from the Relativized Quark Model \cite{Godfrey:1985xj}, the numbers of which we take from \cite{Godfrey:2016nwn}. To
  translate our result for the binding energy to the result displayed in this  plot a isosymmetric kaon mass of $494.6$~MeV has been used. The magenta dotted lines are meant for a better comparison of our results to previous predictions. Another recent Lattice QCD result without uncertainty quantification can be found in \cite{Gayer:2024akw}}
    \label{fig:bs_review}
\end{figure*}

We have applied our nonperturbative tuning methodology to the bottom quark, following the procedure we previously used for the charm \cite{Hudspith:2021iqu}, tuning now to low-lying bottomonia. We find large values of the two parameters $r_s$ and $\nu$ that are proportional to the heavy quark mass. We observe that there is a persistent hierarchy $c_B>c_E$ and that both are close to the nonperturbative $c_{SW}$. We note that all of the parameters of our chosen RHQ action have some interdependence, depending strongly on the choice of lattice spacing. Incorporating our action into OpenQCD was challenging and the cheapest way we found to determine bottom quark propagators is by a fixed number of iterations of the Hopping Parameter Expansion.

As an important test of our nonperturbative tuning of the b-quark RHQ action and to get confidence in applying our setup to the spectroscopy of exotic-mesons in the bottom-strange spectrum, we determined the 1S hyperfine splittings of $B$ and $B_s$ mesons. After taking the continuum limit, we find our resulting postdictions of these benchmark quantities to be consistent with the respective values reported by the Particle Data Group \cite{ParticleDataGroup:2024cfk}. This is a strong test of the suitability for our action in the B-meson sector, giving us confidence to address the largest systematic uncertainty of our previous Lattice-NRQCD study \cite{Hudspith:2023loy}.

Our main physics objective in this work was to arrive at precise predictions for the exotic $B_{s0}^*$ and $B_{s1}$ mesons, which we (and others) have previously determined to lie below the close-by two-meson thresholds, $BK$ and $B^*K$ respectively. We have illustrated, for the first time, that these states persist after the continuum limit has been taken, and are relatively deeply bound with respect to their nearest respective two-meson thresholds $BK$ and $B^*K$. Our final results for the binding energies are:
\begin{equation}
\begin{aligned}
\Delta_{B_{s0}^*} &= -65.9(6.0)(3.0)_\text{Iso} \text{ MeV},\\
\Delta_{B_{s1}} &= -60.6(6.6)(3.0)_\text{Iso}(1.0)_\text{GEVP} \text{ MeV},
\end{aligned}
\end{equation}
with the ``Iso" systematic stemming from the the difference between the isosymmetric kaon mass of 494.6~MeV we used and the (heavier) neutral kaon. The final central values are reasonably consistent with what we predicted from our previous Lattice-NRQCD study, where we had to inflate our systematic uncertainty considerably due to the lack of a formal continuum limit. In comparison to the results presented here, we previously estimated slightly deeper binding. This is a good indication of the difficulty in estimating systematics from Lattice-NRQCD and why a ``continuum limit" should never be attempted when using Lattice-NRQCD data.

Figure \ref{fig:bs_review} shows a comparison of the $B_{s0}^*$ and $B_{s1}$ masses from various models or EFT+model calculations as well as from Lattice QCD calculations \cite{Bardeen:2003kt,Guo:2006fu, Guo:2006rp,Badalian:2007yr,Colangelo:2012xi,Altenbuchinger:2013vwa,Wang:2015mxa,Du:2017zvv,Alhakami:2020vil,Guo:2021rjv,Fu:2021wde,Koponen:2007nr,Gregory:2010gm,Wurtz:2015mqa,Lang:2015hza,Hudspith:2023loy}. In this plot we restrict our comparison to published results which provide at least a partial determination or estimate of the uncertainties. With the exception of early lattice QCD calculations, predictions are mostly in agreement with each other, although the level of rigor of the determination of the associated uncertainties and the r\^ole of phenomenological input differs greatly between the different determinations.

In comparison to previous Lattice QCD studies, this work is the first that takes the continuum limit in a controlled manner using 5 lattice spacings to do so. Previous calculations used techniques where this limit is ill-defined (Lattice-NRQCD) or worked at a single lattice spacing and had to estimate the continuum result and apply a corresponding systematic. Appropriately taking a continuum limit is a crucial step forward in reducing the systematic uncertainty of predictions of these exotic $B_s$ states from Lattice QCD. Further refinements are possible in the future; it would be desirable to determine the finite-volume spectrum with a larger interpolator basis and with a large set of gauge-field ensembles along multiple of the CLS quark-mass trajectories, as well as determining the associated $B^{*}K$ or coupled-channel scattering amplitudes using L\"uscher's finite volume method \cite{Luscher:1985dn,Luscher:1990ux}.

Overall, recent Lattice QCD results combined with phenomenological predictions and the well-established spectrum in the charm sector provide clear indications for the existence of close-to-threshold bound states with quantum numbers of the missing $B_{s0}^*$ and $B_{s1}$. These can be seen as exotic in the sense that they are not in accordance with simple quark-models. They emerge in state-of-the art chiral effective field theory and Lattice QCD calculations alike, which demonstrates the power of these methods to calculate properties of hadrons from QCD. It is likely that both of these states will be determined experimentally in the near future.

\begin{acknowledgments}
The authors would like to thank Fernando Alvarado, Barbara Cid~Mora, Feng-Kun~Guo, Christoph Hanhart, Matthias~F.M.~Lutz, Konstantin Ottnad, Luka Leskovec, Sa\v{s}a Prelov\v{s}ek, and David Thoma for useful discussions. We also would like to thank Marco Pappagallo and Liupan An for alerting us of the preliminary results from LHCb during the drafting of this manuscript. RJH is supported by the U.S. National Science Foundation (NSF) under grant OAC-231143.
DM is funded by the Heisenberg Programme of the Deutsche Forschungsgemeinschaft (DFG, German Research Foundation) project number 454605793.
Calculations for Tuning runs were partly performed on the HPC cluster “Mogon II” at JGU Mainz. This research was supported in part by the cluster computing resource provided by the IT Division at the GSI Helmholtzzentrum f\"ur Schwerionenforschung, Darmstadt, Germany (HPC cluster Virgo). This research used resources of the National Energy Research Scientific Computing Center (NERSC), a Department of Energy User Facility. For the neural network training and predictions we made use of the Keras API. For the light, strange, and bottom quark propagator inversions we used a modified version of the package OpenQCD v1.6\cite{Luscher:2012av}. We thank our colleagues in the CLS consortium for sharing ensembles.
\end{acknowledgments}

\appendix

\vspace{-12pt}
\section{HPE\label{app:hopping}}
\vspace{-4pt}

We originally implemented our RHQ action \cite{Aoki:2003dg} for the charm sector \cite{Hudspith:2021iqu} within the library \verb|openQCD| \cite{Luscher:2012av}. However, we found that none of the solver algorithms could handle the heavy masses for our bottom-quark propagator inversions. All of the solver algorithms available rely on computing the true (explicit) residual for the stopping condition of their mixed-precision solves, which suffers from substantial round-off errors for very heavy quarks. Our correlators can span more than 75 orders of magnitude and the true residual calculation will reach double precision and stop in a handful of iterations, providing the first few time positions away from the source. Clearly a more robust stopping condition was needed as even moving to quad precision would not have alleviated this problem.

\begin{figure}[tb]
\includegraphics[scale=0.28]{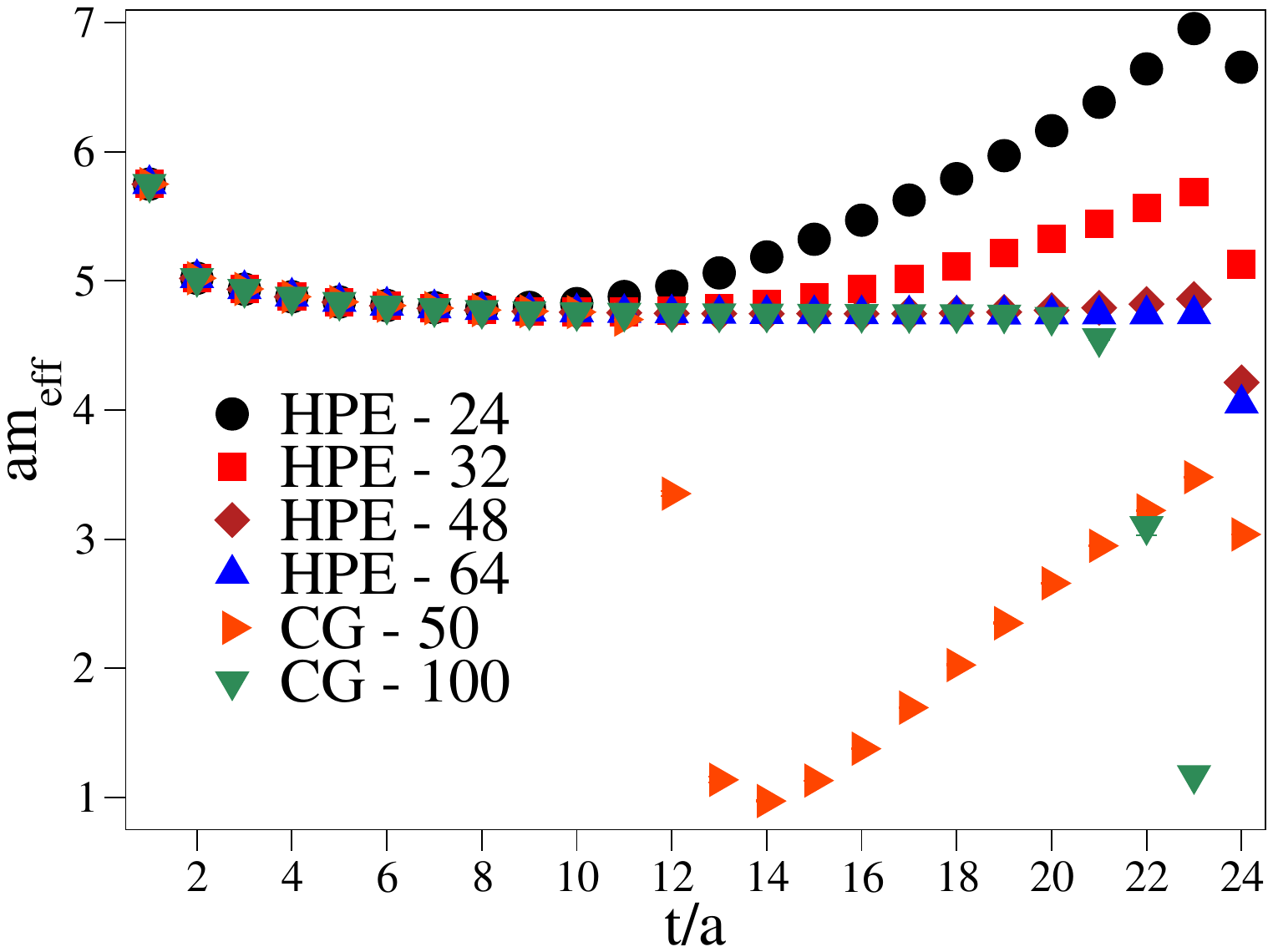}
\caption{Plot of the convergence of fixed-iteration CGNE and the HPE for obtaining our $\eta_b$ mass.}\label{fig:CGHPE}
\end{figure}

At first we attempted the distance preconditioning of \cite{deDivitiis:2010ya}, but found it was too costly as it turns a heavy quark problem into a somewhat ill-conditioned light-quark one. Our second attempt was to use a fixed number of iterations of the Conjugate Gradient algorithm, monitoring the convergence of the effective mass of the $\eta_b$ with increasing number of CG iterations until it stabilised. After our tuning of RHQ parameters, for our $B$ and $B_s$ meson data production, we moved to using a fixed number of iterations of the HPE \cite{Henty:1992cw} as this gives more than a factor of 2 speedup in computing RHQ bottom propagators compared to the fixed-iteration CG for comparable convergence. A comparison of the two methods (CG vs HPE) for the $\eta_b$ can be seen in Fig.~\ref{fig:CGHPE}.

Generically a local fermion action's (for example here we will consider first the Wilson action) propagator $\psi$ prepared from some source $\eta$
\begin{equation}
(1-\kappa\slashed{D})\psi = \eta
\end{equation}
can be inverted by a naive Taylor expansion (Hopping Parameter Expansion):
\begin{equation}
\psi = \eta + \sum_{i=1}^{\infty} (\kappa \slashed{D})^i\eta.
\end{equation}
Or written via a recursion \cite{Henty:1992cw}, where i indicates the i-th iteration:
\begin{equation}
\begin{aligned}
\psi^0 &= \eta,\\
\psi^i &= \psi^{i-1} + \kappa \slashed{D}\psi^{i-1}\;.
\end{aligned}
\end{equation}

We will use even-odd preconditioned clover-improved Wilson fermions so the Taylor expansion is a little more complicated but can be written as:
\begin{equation}
\begin{aligned}
\psi^0 &= A^{-1} \eta,\\
\psi^i &= \psi^{i-1} + (-1)^i ( A_{oo}^{-1} \slashed{D}_{oe}\psi_e^{i-1} + A_{ee}^{-1}\slashed{D}_{eo}\psi_o^{i-1})\;,
\end{aligned}
\end{equation}
where $A^{-1}$ is the inverse of the $c_{SW}$ term.

We will want to truncate the series/recursion at some $n_\text{max}$ where the $\eta_b$ correlator has suitably converged. As our $\kappa$ is often very small such a series should converge rapidly, however for charm and lighter quarks there are much better inversion techniques available.

In comparison to say the Conjugate Gradient on the normal equations CGNE, the HPE per iteration calls either the eo and oe Dirac operators alternatingly, instead of both being called per iteration for the CGNE. We observe that the convergence of the HPE is faster for our parameters. An example of convergence with the HPE order for our coarsest ensemble (A653) for the effective mass of the $\eta_b$ is shown in Fig.~\ref{fig:CGHPE}. As one can see 50 iterations of the CGNE has more contamination than 48 iterations of the HPE while being a factor of 2 more expensive computationally. 

Ultimately there is a trade-off; as the lattice spacing decreases, $\kappa_b$ necessarily grows and the need for higher iterations of the HPE does too. Likewise, the condition number naturally worsens and more CGNE iterations would also be required. As we use an expensive heavy-quark action, which has twice the arithmetic intensity of the light-quark action in the application of the Dirac operator, a single RHQ b-quark inversion often lies somwehere between one and two times the cost of a vanilla strange-quark inversion.

\vspace{-12pt}
\section{GEVP analysis for \texorpdfstring{$B_{s0}^*$}{Bs0*} and \texorpdfstring{$B_{s1}$}{Bs1}}
\label{app:gevp}
\vspace{-4pt}

We investigate possible contamination of our ground-state determination for the $B_{s1}$ by forming an asymmetric $2\times 2$ GEVP (for the exemplary ensemble B450, with reduced statistics) with the additional tensor-like operator ``$T_{ij}$":
\begin{equation}
O_{B_{s1}^\prime,i} = \epsilon_{ijk}\bar{b} \gamma_j \gamma_k s.
\end{equation}
This has the same $J^P=1^+$ quantum numbers as our usual axial vector ``$A_{i}$" operator in Eq.~\ref{eq:exotic_ops}.

\begin{figure}[tb]
\includegraphics[scale=0.26]{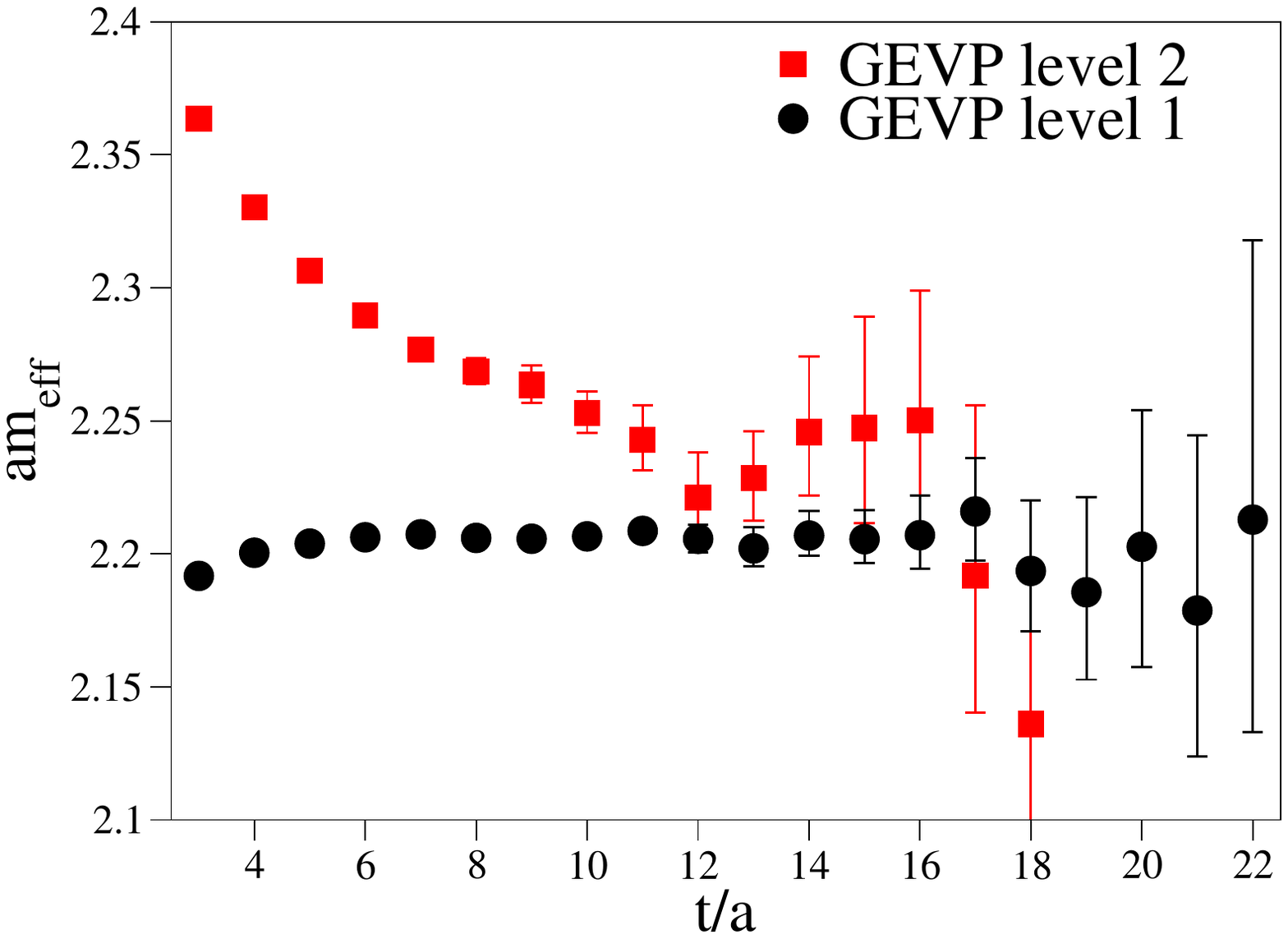}
\caption{Effective mass of the two states from our GEVP for the $B_{s1}$ on ensemble B450.}\label{fig:axeffmass_gevp}
\end{figure}

\begin{figure}[tb]
\includegraphics[scale=0.26]{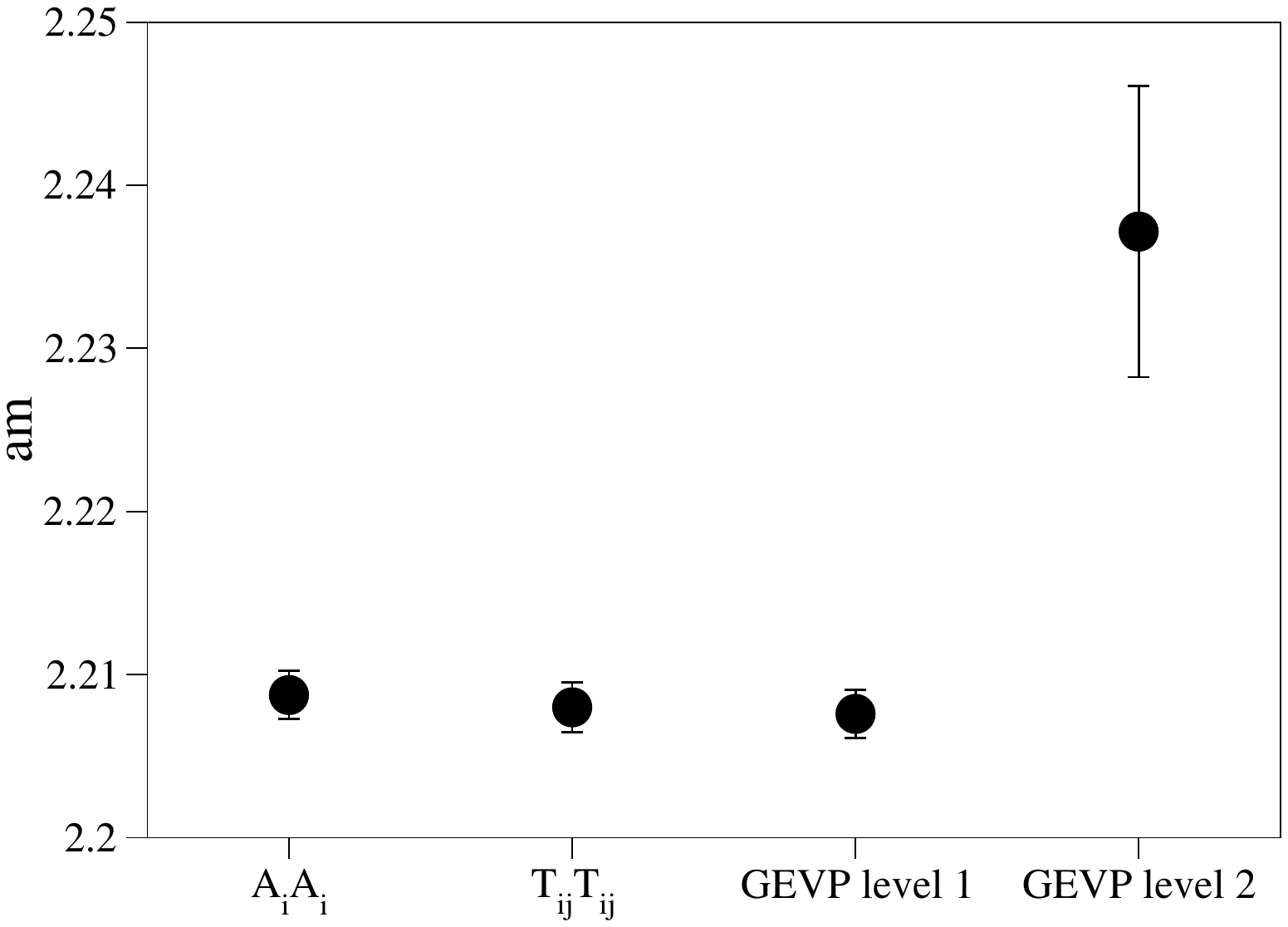}
\caption{Determined energies from our two operators individually or from the GEVP.}\label{fig:axgev_energies}
\end{figure}

Figure~\ref{fig:axeffmass_gevp} shows that the $2\times 2$ GEVP gives two different states, with the excited state being considerably noisier and taking longer to plateau. Figure~\ref{fig:axgev_energies} shows the masses from a fit to either of our operators individually or the two energies fit from the GEVP in the plateau region. We see that we could use either the ``$A_i$" or ``$T_{ij}$" operator for the determination of the ground state, or the first level of a 2x2 GEVP, which in this example would give a slightly lower mass, i.e. an $\mathcal{O}$(1 MeV) deeper binding compared to the lowest two-meson threshold. 

We find the splitting between the two levels to be 76(23) MeV - comparable to the splitting between the $D_{s1}(2536)$ and the $D_{s1}(2460)$ in the charm sector, and compatible with the splitting between the two $J^P=1^+$ states previously determined in \cite{Lang:2015hza}.
Following the arguments for s-wave bound states in \cite{Sasaki:2006jn}, we expect further close-by energy levels from the presence of finite-volume hadron-hadron scattering states should appear above the non-interacting threshold in our situation, making the identification of the level close to threshold with the second quark-model like $D_{s1}$ state likely. Without a larger basis of operators, including scattering operators, not too much faith should be wrested from these results.

\vspace{-12pt}

\begin{table*}
\centering
\begin{tabular}{c|c|c|c|c|c}
\toprule
Ensemble & $m_{B^*}-m_B$ & $m_{B_s^*}-m_{B_s}$ & $m_{B_s}-m_B$ & $m_{B_{s0}^*}-(m_B+m_K)$ & $m_{B_{s1}}-(m_{B^*}+m_K)$ \\
\hline
\multirow{2}{*}{A653} & 30.4(0.4) & 30.4(0.4) & \multirow{2}{*}{0} & \multirow{2}{*}{-57.4(3.1)} & \multirow{2}{*}{-61.4(3.6)} \\
                      & 30.7(0.4)     & 30.7(0.4)     & & & \\
\multirow{2}{*}{A654} & 30.3(0.5) & 31.0(0.4) & \multirow{2}{*}{33.2(0.6)} & \multirow{2}{*}{-57.8(3.9)} & \multirow{2}{*}{-63.1(3.8)} \\
                      & 30.6(0.5)     & 30.7(0.5)     & & & \\
\multirow{2}{*}{GSI\_B650} & 28.4(1.0) & 30.7(0.5) & \multirow{2}{*}{55.3(1.3)} & \multirow{2}{*}{-45.4(4.5)} & \multirow{2}{*}{-49.2(5.5)} \\
                      & 29.0(0.7)     & 30.9(0.5)     & & & \\
\hline
\multirow{2}{*}{U103} & 34.3(2.2) & 34.3(2.2) & \multirow{2}{*}{0} & \multirow{2}{*}{-68.0(7.3)} & \multirow{2}{*}{-75.8(9.4)} \\
                      & 34.9(0.8)     & 34.9(0.8)     & & & \\
\multirow{2}{*}{H101} & 34.0(1.5) & 34.0(1.5) & \multirow{2}{*}{0} & \multirow{2}{*}{-57.5(7.4)} & \multirow{2}{*}{-57.8(7.0)} \\
                      & 33.3(0.9)     & 33.3(0.9)     & & & \\
\multirow{2}{*}{H102} & 34.3(0.8) & 34.3(0.7) & \multirow{2}{*}{23.4(0.5)} & \multirow{2}{*}{-52.7(4.1)} & \multirow{2}{*}{-55.8(5.8)} \\
                      & 34.2(0.7)     & 34.6(0.6)     & & & \\
\multirow{2}{*}{H105} & 31.5(2.5) & 34.7(0.6) & \multirow{2}{*}{50.8(1.9)} & \multirow{2}{*}{-48.0(4.0)} & \multirow{2}{*}{-48.8(6.2)} \\
                      & 33.6(0.8)     & 34.9(0.6)     & & & \\
\multirow{2}{*}{X150} & 32.7(0.7) & 33.3(0.5) & \multirow{2}{*}{50.4(0.9)} & \multirow{2}{*}{-50.7(2.7)} & \multirow{2}{*}{-50.2(2.8)} \\
                      & 32.8(0.7)     & 33.3(0.5)     & & & \\
\multirow{2}{*}{N101} & 34.6(1.5) & 35.0(0.7) & \multirow{2}{*}{52.6(1.7)} & \multirow{2}{*}{-46.0(4.4)} & \multirow{2}{*}{-49.9(5.9)} \\
                      & 34.7(0.7)     & 35.4(0.6)     & & & \\
\hline
\multirow{2}{*}{B450} & 36.8(0.8) & 36.8(0.8) & \multirow{2}{*}{0} & \multirow{2}{*}{-66.1(6.2)} & \multirow{2}{*}{-67.0(6.2)} \\
                      & 37.3(0.7)     & 37.3(0.7)     & & & \\
\multirow{2}{*}{X451} & 36.5(0.7) & 37.7(0.6) & \multirow{2}{*}{48.1(0.9)} & \multirow{2}{*}{-53.0(3.4)} & \multirow{2}{*}{-54.0(4.4)} \\
                      & 36.6(0.7)     & 37.7(0.6)     & & & \\
\hline
\multirow{2}{*}{H200} & 42.3(1.7) & 42.3(1.7) & \multirow{2}{*}{0} & \multirow{2}{*}{-79.6(9.6)} & \multirow{2}{*}{-80.8(11.5)} \\
                      & 40.4(1.0)     & 40.4(1.0)     & & & \\
\multirow{2}{*}{N202} & 38.1(1.6) & 38.1(1.6) & \multirow{2}{*}{0} & \multirow{2}{*}{-61.6(4.9)} & \multirow{2}{*}{-57.3(5.7)} \\
                      & 39.5(1.2)     & 39.5(1.2)     & & & \\
\multirow{2}{*}{N203} & 40.3(1.0) & 40.5(0.8) & \multirow{2}{*}{27.4(0.6)} & \multirow{2}{*}{-58.3(4.0)} & \multirow{2}{*}{-59.2(4.5)} \\
                      & 40.0(0.8)     & 40.7(0.7)     & & & \\
\multirow{2}{*}{N200} & 37.2(1.4) & 38.6(1.0) & \multirow{2}{*}{48.5(1.4)} & \multirow{2}{*}{-54.2(4.9)} & \multirow{2}{*}{-48.9(5.4)} \\
                      & 38.7(1.0)     & 40.5(0.8)     & & & \\
\multirow{2}{*}{D200} & 40.6(1.2) & 41.7(0.9) & \multirow{2}{*}{70.8(2.4)} & \multirow{2}{*}{-59.1(4.3)} & \multirow{2}{*}{-59.0(5.5)} \\
                      & 40.3(0.9)     & 41.4(0.7)     & & & \\
\hline
\multirow{2}{*}{N300} & 44.6(1.3) & 44.6(1.3) & \multirow{2}{*}{0} & \multirow{2}{*}{-67.5(5.7)} & \multirow{2}{*}{-68.0(6.9)} \\
                      & 45.7(1.0)     & 45.7(1.0)     & & & \\
\botrule
\end{tabular}
\caption{Table of results for our various $B$- and $B_s$-meson splittings, all results are given in MeV. The upper parts for the $B^*-B$ and $B_s^*-B_s$ are from the non-ratio analysis and the lower are from the ratio.}\label{tab:all_results}
\end{table*}

\section{Table of results}\label{app:results}

Table \ref{tab:all_results} lists our results for the ensembles given in Tab.~\ref{tab:configs}.

\clearpage

\bibliographystyle{apsrev4-2}
\bibliography{RHQ}

\end{document}